\documentclass[11pt]{article}
\pdfoutput=1
\newcommand{\Comment}[1]{{}}
\usepackage{mathtools}
\usepackage{amsfonts,amsthm,amsmath,amssymb,slashed}
\usepackage[textwidth = 430 pt, textheight = 630 pt]{geometry}
\usepackage{color}
\usepackage{booktabs}
\usepackage{subcaption}
\usepackage[export]{adjustbox}

\Comment{\usepackage{color}
\definecolor{MyDarkBlue}{rgb}{0.15,0.15,0.45}
\usepackage[linktocpage=true]{hyperref}
\hypersetup{
colorlinks=true,
citecolor=MyDarkBlue,
linkcolor=MyDarkBlue,
urlcolor=MyDarkBlue,
pdfauthor={Dmitry Melnikov and Horatiu Nastase},
pdftitle={Wiedemann-Franz laws and $Sl(2,\mathbb{Z})$ duality in AdS/CMT holographic duals and one-dimensional effective actions for them},
pdfsubject={hep-th}
}
\usepackage[utf8]{inputenc}
\usepackage[numbers,sort&compress]{natbib}
\usepackage{hypernat}}
\usepackage{graphicx}
\usepackage{cite}
\usepackage[colorlinks=true]{hyperref}
\hypersetup{
  allcolors = blue,
}

\newcommand\ignore[1]{}
\def\one{{\,\hbox{1\kern-.8mm l}}}

\def\a{\alpha}\def\b{\beta}

\def\d{\partial}

\newcommand{\Cset}{{\,\,{{{^{_{\pmb{\mid}}}}\kern-.45em{\mathrm C}}}}}

\newcommand{\be}{\begin{equation}}
\newcommand{\bea}{\begin{eqnarray}}

\newcommand{\ee}{\end{equation}}
\newcommand{\eea}{\end{eqnarray}}

\begin{document}

\renewcommand{\thefootnote}{\fnsymbol{footnote}}

\makeatletter
\@addtoreset{equation}{section}
\makeatother
\renewcommand{\theequation}{\thesection.\arabic{equation}}

\rightline{}
\rightline{}
%   \vspace{1.8truecm}

\begin{flushright}
ITEP-TH-28/20
% preprint nrs.
\end{flushright}

\vspace{10pt}

%%%%%%%%%%%%%%%%%

\begin{center}
{\LARGE \bf{\sc Wiedemann-Franz laws and $Sl(2,\mathbb{Z})$ duality in AdS/CMT holographic duals and one-dimensional effective actions for them}}
\end{center}
 \vspace{1truecm}
\thispagestyle{empty} \centerline{
{\large \bf {\sc Dmitry Melnikov${}^{a,b}$}}\footnote{E-mail address: \Comment{\href{mailto:dmitry@iip.ufrn.br}}{\tt dmitry@iip.ufrn.br}}
{\bf{\sc and}}
{\large \bf {\sc Horatiu Nastase${}^{c}$}}\footnote{E-mail address: \Comment{\href{mailto:horatiu.nastase@unesp.br}}{\tt horatiu.nastase@unesp.br}}
                                                        }

\vspace{.5cm}

\centerline{{\it ${}^a$International Institute of Physics, Universidade Federal do Rio Grande do Norte, }}
\centerline{{\it Campus Universit\'{a}rio, Lagoa Nova, Natal-RN 59078-970, Brazil}}

\vspace{.3cm}

\centerline{{\it ${}^b$Institute for Theoretical and Experimental Physics,  }}
\centerline{{\it B. Cheremushkinskaya 25,  Moscow 117218,  Russia}}

\vspace{.3cm}

\centerline{{\it ${}^c$Instituto de F\'{i}sica Te\'{o}rica, UNESP-Universidade Estadual Paulista}}
\centerline{{\it R. Dr. Bento T. Ferraz 271, Bl. II, Sao Paulo 01140-070, SP, Brazil}}

\vspace{1truecm}

%%%%%%%%%%%%%%%%%
\thispagestyle{empty}

\centerline{\sc Abstract}

\vspace{.4truecm}

\begin{center}
\begin{minipage}[c]{380pt}
{\noindent In this paper we study the Wiedemann-Franz laws for transport in 2+1 dimensions, and the action 
of $Sl(2,\mathbb{Z})$ on this transport, for
theories with an AdS/CMT dual. We find that $Sl(2,\mathbb{Z})$ restricts the 
RG-like flow of conductivities and that the Wiedemann-Franz law is $\bar L 
=\bar\kappa/(T\sigma)=cg_4^2\pi/3$, from the weakly coupled} gravity dual. 
 In a self-dual theory this value is also the value of $L 
=\kappa/(T\sigma)$ in the weakly coupled field theory description. Using the 
formalism of a 0+1 dimensional effective action for both generalized $SYK_q$ 
models and the $AdS_4$ gravity dual, we calculate the transport coefficients and 
show how they can be matched at large $q$. We construct a generalization 
of this effective action that is invariant under $Sl(2,\mathbb{Z})$  and 
can describe vortex conduction and integer quantum Hall effect.
%including both magnetic conduction and an integer Hall conductivity.

\end{minipage}
\end{center}

\vspace{.5cm}

\setcounter{page}{0}
\setcounter{tocdepth}{2}

\newpage

%\tableofcontents
\renewcommand{\thefootnote}{\arabic{footnote}}
\setcounter{footnote}{0}

\linespread{1.1}
\parskip 4pt

%{}~
%{}~

%---------------------------------------------------------

%%%%%%%%%%%%%%%%%%%%%%%%%%%%%%%%%%%%%%%%%%%%%%%%%%%%%%%%%%%%%%%%%%%%%%%%%%%%%%%%%%%%%%%%
\section{Introduction}
%%%%%%%%%%%%%%%%%%%%%%%%%%%%%%%%%%%%%%%%%%%%%%%%%%%%%%%%%%%%%%%%%%%%%%%%%%%%%%%%%%%%%%%%

In classical Fermi liquid theory in 2+1 dimensions, one obtains the Wiedemann-Franz 
law for the ratio of the off-diagonal heat transport coefficient 
$\kappa_{xy}/T$ to the Hall conductivity in the $T\rightarrow 0$ limit, 
\be
\label{WF1}
\frac{\kappa_{xy}/T}{\sigma_{xy}}\rightarrow 
\frac{\pi^2}{3}\left(\frac{k_B}{e}\right)^2.
\ee

However, it was known since the work of Kane and Fisher \cite{Kane:1997fda} that the right-hand side, called the Lorenz number $L$, can 
in general be multiplied by some object, which later, in the work of Read and Green for the Fractional Quantum Hall Effect (FQHE)
\cite{Read:1999fn} was identified with a central charge $c$. 
But in such more complicated systems, like systems described by conformal Abelian Chern-Simons analyzed in \cite{Witten:2003ya}, 
there is an action of an $ Sl(2,\mathbb{Z})$ symmetry on them, including a $T$ and an $S$ generator. The $S$ generator acts 
via the particle-vortex symmetry, which can be defined as in terms of field theory in \cite{Burgess:2000kj}, and better defined in 
\cite{Murugan:2014sfa} (at the level of the path integral), and results in an action on the complex electrical conductivity $\sigma=\sigma_{xy}
+i\sigma_{xx}$.

It is then of interest to see what we can obtain obtain for the Wiedemann-Franz 
law and $ Sl(2,\mathbb{Z})$ symmetry in systems described holographically, via 
the AdS/CMT correspondence (see for instance the book \cite{Nastase:2017cxp} for 
a review).  Transport in 2+1 dimensional case was discussed in a number 
of papers: the first results for systems in magnetic field were obtained in 
\cite{Hartnoll:2007ai,Hartnoll:2007ih}. In \cite{Hartnoll:2007ih}, the 
symmetry of the charge and heat transport coefficients with respect to the 
electric-magnetic duality action was observed in both the holographic and 
magnetohydronamic approaches. One of the questions that we will address here is 
the relations of those results with the above classical Wiedemann-Franz law, 
with $L=\pi^2/3$, as well as with the non-Fermi liquid extension, with a 
nontrivial central charge $c$ coefficient.

We will also consider a slightly more general holographic setup describing 
a \emph{disordered} system in the presence of a nontrivial electric charge 
density $\rho$ and magnetic field $B$, as well as a nontrivial $\theta 
F_{\mu\nu}\tilde F^{\mu\nu}$ term, corresponding to Chern-Simons in 3 
dimensions, as in  \cite{Alejo:2019utd}
(following \cite{Alejo:2019hnb} and \cite{Blake:2015ina}). We will analyze the 
Lorenz number $L$ coming from that calculation, as well as the analogous law for 
the dissipative components $xx$, and the action of the $Sl(2,\mathbb{Z})$ on 
the transport coefficients and the laws.  We will show that the dependence 
of the conductivities from the disorder in the holographic model match 
precisely the hydrodynamic analysis of \cite{Hartnoll:2007ih} if the 
rescaled disorder parameter is replaced by the inverse scattering time. 
Even though we work with DC conductivities, the frequency dependence can be 
obtained by a shift of the disorder parameter.

We show that the coupling constants of the 3+1 dimensional gravity Lagrangian 
can be thought of as bare values of the conductivity, as in effective two 
dimensional sigma models of conductivity in disordered systems. Magnetic field 
is a relevant perturbation, which drives the system to a IR fixed 
point, corresponding to the quantum Hall regime, with zero direct conductivity. 
The electric-magnetic duality restricts this flow to the fundamental domain of 
$Sl(2,\mathbb{Z})$. Integer shifts of the theta term translate the fundamental 
domain in the upper half plane.

As far as the Wiedmann-Franz law is concerned we find that at the fixed point 
with no disorder the most interesting quantity is not the thermal conductivity 
$\kappa$, but rather the heat conductivity $\bar\kappa$, so that at low 
temperature one gets the modified Lorenz number
\be
\bar{L} \ \equiv \ \frac{\bar\kappa/T}{\sigma} \ = \ \frac{\pi}{3}\,cg_4^2 + 
O(T)\,.
\ee
in terms of the centeral charge $c$ and the gravity coupling $g_4$. In the 
normalization of \cite{Hartnoll:2007ih} $cg_4^2=\pi$ so $\bar L$ takes the 
value of the classical Lorenz number~(\ref{WF1}). The interesting part is that this specific normalization 
is the self-dual point of the model, according to~\cite{Herzog:2007ij}. Since the duality exchanges 
$\kappa_{xy}\leftrightarrow \bar\kappa_{xy}$, $\bar L$ calculated in the gravity dual actually 
measures $L$ in the weakly coupled gauge theory. 
This feature does not hold for a generic holographic model, and gravity is not expected to capture 
the weak coupling regime of the dual theory. However, if self-duality is exact, it should provide a window into weak coupling.

The strong coupling value of the Lorenz number is
\be
L \ \equiv \ \frac{\pi}{3}\frac{c}{g_4^2}\frac{1}{\sigma_{xy}^2} + O(T)\,.
\ee

At the self-dual point it is equal to $c^2/(3\sigma_{xy}^2)$.

In the case of direct conductivities, similar result holds in the limit of zero 
magnetic field, and then zero disorder, in which case the modified Lorenz 
number has a finite value ${\bar L}_{xx}=\pi/3(cg_4^2)$. This is again mapped 
to the weak coupling value by the duality.

Further, in \cite{Davison:2016ngz}, the Wiedemann-Franz law was analyzed from 
the point of view of $SYK_q$ generalizations of the SYK model for $q$ fermion 
interactions, both in $d$ dimensional lattices of $SYK_q$ models, and in the 
gravity duals corresponding to the same physics.  For the special 
choice of intersite coupling the Lorenz number of the $SYK_q$ model was 
shown to be given by
\be
L \ = \ \frac{4\pi^2}{3q^2}\,,
\ee
with $q=2$ corresponding to the free theory point. However, this result was not 
reproduced by the gravity calculations.

A 0+1 dimensional effective action generalizing the Schwarzian action for the 
standard SYK model was also found in \cite{Davison:2016ngz} for complex 
fermions with charge. We will show that it can be used to directly find the 
transport in the charged $SYK_q$ models. We will also show how it can 
be matched to the holographic calculation in the large $q$ 
limit.  We will find that the ratio of the heat and charge 
susceptibilities is also expressed as
\be
\frac{\gamma}{K} \ = \ \frac{\pi}{3}c g_4^2\,,
\ee
at zero magnetic field. Moreover, we will show that we can 
extend the 0+1 dimensional effective 
action to one that is selfdual under electric-magnetic S duality, and we can 
also add a $\theta$ term corresponding to the T operation, thus arriving at an 
$Sl(2,\mathbb{Z})$ invariant form.

The paper is organized as follows. In section~\ref{sec:WFlaw} we define the 
transport coefficients, and describe the general expectations for the 
Wiedemann-Franz laws. In section~\ref{sec:SL2Zandtransport} we first find the 
holographic transport coefficients and W-F laws, then describe the effect of 
$Sl(2,\mathbb{Z})$ on them, and finally generalize the calculation of 
\cite{Hartnoll:2007ih} to the case of nontrivial central charge $c$. In 
section~\ref{sec:1dTranportCoeffs} we calculate the transport coefficients from 
the 0+1 dimensional generalized Schwarzian effective action, and compare with 
the holographic calculations. In section~\ref{sec:1dEffAct} we write the 
self-dual version of the 0+1 dimensional effective action, and find the $\theta$ 
term for the same, and in section~\ref{sec:conclusions} we conclude. In 
Appendix~\ref{sec:app1} we review the action of particle-vortex duality, and in 
Appendix~\ref{SYKApp} -- an alternative calculation of conductivities from the 
effective action of the generalized SYK models.

\section{Wiedemann-Franz laws in condensed matter and general theory}
\label{sec:WFlaw}

\subsection{Definition of transport coefficients in various dimensions}

In this paper, we will be interested in electric and heat transport, and the corresponding transport coefficients. 

Transport coefficients are defined as the coefficients for the linear response (electric current $\vec{J}$, heat current $\vec{Q}$, etc.) 
of the material to external fields: external electric field $\vec{E}$ and 
external temperature gradient $\vec{\nabla}T$ in our case. One can write then (in the convention in  \cite{Hartnoll:2007ih}) the matriceal relation 
\be
\begin{pmatrix}\vec{J}\\ \vec{Q}\end{pmatrix} = \begin{pmatrix} \hat \sigma & \hat \a\\ T\hat \a &\hat {\bar \kappa}
\end{pmatrix} \begin{pmatrix} \vec{E}\\ -\vec{\nabla}T\end{pmatrix}\;,
\ee
where $\hat \sigma$ is the matrix of electric conductivities, $\hat \a$ is the matrix of thermoelectric conductivites
and the matrix of thermal conductivities is 
\be
\label{heatvstherm}
\hat \kappa=\hat{\bar \kappa} -T\hat \a \hat \sigma^{-1}\hat \a.
\ee

One can also define the thermoelectric power coefficient 
\be
\hat \theta=-\hat \sigma^{-1}\hat \a\;,
\ee
so that the Nernst coefficient is 
\be
\nu=\frac{\theta_{xy}}{B}.
\ee

Note that instead of using matrices, {\em in the specific case of isotropic 
coefficients, i.e., that $\sigma_{xx}=\sigma_{yy}$
and $\sigma_{xy}=-\sigma_{yx}$ and similar for all the other transport coefficients}, so
\be
\sigma=\sigma_{xx}\one+\sigma_{xy}{\bf \epsilon}\;,\;\;
\a=\a_{xx}\one+\a_{xy}{\bf \epsilon}\;,\;\;  
\kappa=\kappa_{xx}\one+\kappa_{xy}{\bf \epsilon}\;,
\ee
we can use complex quantities, 
\be
\sigma\equiv \sigma_{xy}+i\sigma_{xx}\;,\;\; A\equiv \a_{xy}+i\a_{xx}\;,\;\; K\equiv \kappa_{xy}+i\kappa_{xx}\;,
\ee
and obtain the same results, in particular
\bea
\frac{\kappa_{xx}}{T}&=&\frac{\bar\kappa_{xx}}{T}-\frac{\a_{xx}^2\sigma_{xx}+2\a_{xx}\a_{xy}\sigma_{xy}-\sigma_{xx}\a_{xy}^2}{
\sigma_{xx}^2+\sigma_{xy}^2}\cr
\frac{\kappa_{xy}}{T}&=&\frac{\bar\kappa_{xy}}{T}-\frac{\a_{xy}^2\sigma_{xy}+2\a_{xx}\a_{xy}\sigma_{xx}-\sigma_{xy}\a_{xx}^2}{
\sigma_{xx}^2+\sigma_{xy}^2}.\label{kappabarkappa}
\eea

One can calculate the above transport coefficients also from the 
diffusivity coefficients $D$ and the susceptibilities $\chi$ in momentum space 
(here 
we follow \cite{Davison:2016ngz} -- one of the original references is 
\cite{Kadanoff:1963}). 
In a general dimension, one defines $\chi_s$ from $\chi$ as 
\be
\chi(k,\omega)=[i\omega(-i\omega +Dk^2)^{-1}+1]\chi_s\;,\label{chichisD}
\ee
where 
\be
\label{diffusivities}
D=\begin{pmatrix} D_1& 0\\ 2\pi {\cal E}(D_1-D_2)& D_2\end{pmatrix}
\ee
and then the matrix of transport coefficients is
\be
\label{Cond2Sus}
\begin{pmatrix} \sigma& \a\\ \a T& \bar \kappa\end{pmatrix}=D\chi_s.
\ee

Note that in the above we can also consider either $\sigma,\a,\bar\kappa$ 
and $D$ to be matrices, or equivalently, to be complex 
objects, with $D=D_{xy}+iD_{xx}$. Since $\chi_s$ are thermodynamical quantities, they are the same for $xx$ and $xy$ components, 
so by taking the real and imaginary parts of the above equation (understood as a complex equation), we obtain 
\be
\begin{pmatrix} \sigma_{xy}& \a_{xy}\\ \a_{xy} T& \bar \kappa_{xy}\end{pmatrix}=D_{xy}\chi_s\;,\;\;\;
\begin{pmatrix} \sigma_{xx}& \a_{xx}\\ \a_{xx} T& \bar \kappa_{xx}\end{pmatrix}=D_{xx}\chi_s.\label{Dcomplex}
\ee

\subsection{General theory expectations and Wiedemann-Franz laws}

In classical condensed matter in 2+1 dimensions, 
Fermi liquid theory obtains the Wiedemann-Franz law for the off-diagonal ("Hall", or Leduc-Righi (LR)) heat transport coefficient
$\kappa_{xy}$ and the Hall electrical conductivity: at temperature $T\rightarrow 0$, 
\be
\frac{\kappa_{xy}/T}{\sigma_{xy}}\rightarrow \frac{\pi^2}{3}.
\ee
More precisely, restoring all dimensions, there is also a $k_B^2$ on the right-hand side, and the corresponding object is called the Lorenz number.

Now, for some materials, the $T\rightarrow 0$ limit of $\kappa_{xy}/\sigma_{xy}$ has a different form. In the (Fractional) Quantum Hall Effect, corresponding to an 
interacting system, the question is {\em why is there a different coefficient in the W-F law, rather than $\pi^2/3$}? Kane and Fisher~\cite{Kane:1997fda} 
have argued that there 
are different Landau levels, meaning different edge modes. There is also the filling fraction appearing both in $\kappa_{xy}$ and in $\sigma_{xy}$, but 
it cancels in the ratio.

More precisely, in the work of Kane and Fisher~\cite{Kane:1997fda}, it is shown that the ratio of the thermal and electric Hall conductivities can be expressed as
\be
\label{KF}
\frac{\kappa_{xy}}{\sigma_{xy}}\ = \  \frac{\sum_a\eta_a}{\sum_a t_a^2\eta_a}\frac{\pi^2}{3}T\;,
\ee
where $\eta_a$ are inverse eigenvalues of an $N\times N$ matrix $K_{ab}$ 
characterizing a given topological order (phase) at the $N$th level of the 
Haldane-Halperin hierarchy. Numbers $t^a$ are charges of elementary 
quasiparticle excitations at this level (for example charges of an electron or a 
hole excitations).  Before diagonalization these numbers are typically assumed 
to be $t_a=1$, which means that possible bound states of electrons are ignored. 
After diagonalization $t_a$ are some numbers depending on $K_{ab}$. In other 
words,
\be
\label{KF2}
\frac{\kappa_{xy}}{\sigma_{xy}}\ = \  \frac{\sum_{ab}\eta_{ab}}{\sum_{ab}K^{-1}_{ab}}\frac{\pi^2}{3}T\;,
\ee

In~\cite{Kane:1997fda} it is somewhat assumed that $\eta_{a}=\pm 1$ (although it 
does not seem to be a generic property~\cite{Wen:1992uk}) so each of these 
numbers represent edge modes moving in the direction prescribed by the magnetic 
field $\eta=1$, or in the opposite direction $\eta=-1$. Each such mode 
contribuites a unit of heat conductivity and $t^2_a$ units of electric 
conductivity, summed algebraically. For example, if all the modes propagate in 
the same direction then $\sum_a\eta_a= N$ counts the number of channels. Also, 
if all $t_a=1$ (after diagonalization), the Lorenz number takes its classical 
value, while its deviation from that value depends on the structure of the 
matrix $K_{ab}$.

Later, in the work of Read and Green \cite{Read:1999fn}, it was argued that in the case of a 2+1 dimensional FQHE system with 2 edges, 
the edge modes control all transport, and we can construct:

-a spin analog of electric Hall conductivity, and obtain, for the p-wave paired state, and with the spin in units of $\hbar$, 
\be
\sigma_{xy}^s=\frac{m}{h}\;,
\ee
where $h$ is the Planck constant (=$2\pi \hbar$) and $m$ is a integer winding number equal to $\pm 1$ in a weak-pairing phase. 
The electric conductivity should have similar properties. Since the electric charge is $e$, we expect the electric conductivity to also be $\sigma_{xy}\propto e^2$. 

-the Hall (or LR) heat conductivity 
\be
\kappa_{xy}=c \frac{\pi^2}{3h}T\;,
\ee
where $c$ is the (Virasoro) central charge of the 1+1 dimensional edge conformal 
field theory (including the edge modes of the FQHE)
 \cite{Bloete:1986qm,Affleck:1986bv,Cappelli:2001mp}. 
In the case that the 2 edge modes move in different directions, the central 
charge should be of the difference between right- and left-moving theories. 
Note however that for a Majorana fermion $c=1/2$, so {\em a priori} $c$ could be half-integer. 

That means that the Lorenz number is proportional to $c$. Note that the central charge $c$ is for the 1+1 dimensional 
edge of the 2+1 dimensional field theory; 
however, if only the edge modes are relevant (as in the case of the FQHE, or for Chern-Simons models), 
it can be thought of as the central charge (or number of degrees of freedom in a generalized sense) for the field theory. 
In dimensions higher than 2, the central 
charge is related to the trace anomaly $\langle {T^\mu}_\mu\rangle$, or anomaly in conformal invariance 
(but is not a central charge in the Virasoro algebra anymore). 

It was also proposed that for the normal (dissipative) 
heat conductivity, 
\be
\kappa_{xx}=\kappa_{yy}\sim \frac{\pi^2}{3h}\frac{T}{g^2}\;,
\ee
where $g$ is the coupling in the two-dimensional nonlinear sigma model 
for disordered electron systems, e.g. \cite{Senthil:1998qu}. 

Moreover, in \cite{Hartnoll:2007ih} it was shown that in fact, in general, and in a holographic model (dyonic black hole in $AdS_4$),  
all transport coefficients come from a universal quantity
\be
\sigma_Q=\frac{4e^2}{h}\Phi_\sigma.
\ee
In particular $\sigma_{xx}=\sigma_Q+...$, and then 
\be
\kappa_{xx}=\frac{\Phi_\sigma}{h} k_B^2T \left(\frac{\epsilon+P}{k_BT\rho}\right)[...]\;,
\ee
where the bracket $[...]$ goes to 1 in a certain limit.

Then we expect that there should be an analog of the Wiedemann-Franz law for the dissipative components in the case of these nontrivial 
strongly coupled systems (FQHE-like), so 
\be
\frac{\kappa_{xx}/T}{\sigma_{xx}}\sim (...)\frac{\pi^2k_B^2}{3}.
\ee

\section{$Sl(2,\mathbb{Z})$ duality on transport coefficients and Wiedemann-Franz law for AdS/CMT holographic models}
\label{sec:SL2Zandtransport}

\subsection{AdS/CMT holographic models and results for transport coefficients}
\label{sec:AdS/CMTandTransport}

AdS/CMT models are usually phenomenological holographic models. To describe 2+1 dimensional matter with charge and heat transport, we need 
to consider a 3+1 dimensional gravitational solution in a phenomenological theory with a gauge field and charged black hole solutions in an AdS background. 

The 3+1 dimensional (holographic bulk) gravitational model contains gravity, 3 scalars and a vector. The action is 
\bea
\label{MasterAction}
S&=&\int d^4x\sqrt{-g} \left[\frac{1}{16\pi G_N}\left(R - 
\frac{1}{2}\left[(\d_\mu\phi)^2 + 
\Phi(\phi)\left((\d_\mu\chi_1)^2+(\d_\mu\chi_2)^2\right)\right]
-V(\phi)\right)\right.\cr
&&\left. -\frac{Z(\phi)}{g_4^2}F_{\mu\nu}^2-W(\phi)F_{\mu\nu}\tilde F^{\mu\nu}\right].
\eea

Note that there are two "axions" $\chi_1,\chi_2$ that are usually ignored, by putting $\Phi(\phi)=0$, but they are 
necessary if we want to have dissipative charge and heat 
transport, since we need to break translational invariance in the $x$ and $y$ directions, achieved 
by having a linear background for the axions. Alternatively, the translational symmetry breaking can 
be introduced via a "holographic lattice", that is considering a metric with explicit dependence of 
spatial coordinates, as for example in~\cite{Donos:2017mhp}. We will not take this path here.

Considering a black hole solution with certain asymptotics of the perturbed AdS 
type in the gravity sector, and then adding boundary sources for the 
perturbation in $\vec{\nabla}T/T$, the electric and magnetic fields 
$\vec{E},\vec{B}$ and charge density $\rho$, we can calculate the transport 
coefficients as follows (see \cite{Iqbal:2008by} for the original idea). We can 
calculate certain fluxes $\vec{\cal J}$ and $\vec{\cal Q}$ (modifications of the 
electric current $\vec{j}$ and heat current $\vec{Q}$) that are $r$-independent, 
so can be calculated at the horizon of the black hole, from the metric 
fluctuations induced at the horizon by the presence of varying sources for 
$\vec{E}$ and $\vec{\nabla}T/T$, in the presence of constant $\vec{B}$ and 
$\rho$. Then the transport coefficients are the linear response coefficients and 
were found to be \cite{Alejo:2019utd}\footnote{See also \cite{Blake:2015ina,Donos:2015bxe,Erdmenger:2016wyp}}
\bea
\sigma_{xx}&=&e^{2V}k^2\Phi \frac{(2\kappa_4^2)\frac{g_4^4}{Z^2}\rho^2+(2\kappa_4^2)B^2+\frac{g_4^2}{Z}e^{2V}
k^2\Phi}{(2\kappa_4^2)^2\frac{g_4^4}{Z^2}B^2\rho^2+(2\kappa_4^2 B^2+\frac{g_4^2}{Z}e^{2V}k^2\Phi)^2}\cr
&=&\frac{e^{2V}k^2\Phi}{(2\kappa_4^2)B^2}\frac{1+\frac{B^2Z^2}{\rho^2g_4^4}
+\frac{Z e^{2V} k^2\Phi}{g_4^22\kappa_4^2\rho^2}}{1+\left(\frac{BZ}{\rho g_4^2}
+\frac{e^{2V}k^2\Phi}{2\kappa_4^2B\rho}\right)^2}\cr
\sigma_{xy}&=&2\kappa_4^2\frac{\rho}{B}\frac{2\kappa_4^2\frac{g_4^4}{Z^2}\rho^2+2\kappa_4^2B^2+2\frac{g_4^2}{Z}
e^{2V}k^2\Phi}{(2\kappa_4^2)^2\frac{g_4^4}{Z^2}B^2\rho^2+(2\kappa_4^2 B^2+\frac{g_4^2}{Z}e^{2V}k^2\Phi)^2}
-4W\cr
&=&\frac{\rho}{B}\frac{1+\frac{B^2Z^2}{\rho^2g_4^4}+2\frac{Z e^{2V} k^2\Phi}{g_4^22\kappa_4^2\rho^2}}{1+\left(\frac{BZ}{\rho g_4^2}
+\frac{e^{2V}k^2\Phi}{2\kappa_4^2B\rho}\right)^2}-4W.\label{chargecond}
\eea
%(note that the $2\kappa_4^2$ factors were kept as in \cite{Alejo:2019utd} for convenience, though they are actually absorbed
%in $k^2\Phi(\phi)$ due to the different normalization for the action for $\chi_{1,2}$ in (\ref{MasterAction})) 
for the electric conductivities,
\bea
\frac{\bar \kappa_{xx}}{T}&=&(2\kappa_4^2)s^2\frac{g_4^2\left[(2\kappa_4^2)B^2Z+
g_4^2e^{2V}k^2\Phi\right]}{(2\kappa_4^2)^2g_4^4\rho^2B^2+((2\kappa_4^2)
B^2Z+g_4^2e^{2V}k^2\Phi)^2}\cr
&=&\frac{s^2}{\rho^2}\frac{\frac{Z}{g_4^2}+\frac{e^{2V}k^2\Phi}{(2\kappa_4^2)B^2}}{1+\left(\frac{BZ}{\rho g_4^2}
+\frac{e^{2V}k^2\Phi}{2\kappa_4^2B\rho}\right)^2}\cr
\frac{\bar \kappa_{xy}}{T}&=&(2\kappa_4^2)s^2\frac{g_4^2(2\kappa_4^2)g_4^2\rho B}{(2\kappa_4^2)^2g_4^4\rho^2B^2+
((2\kappa_4^2)B^2Z+g_4^2e^{2V}k^2\Phi)^2}\cr
&=&\frac{s^2}{\rho B}\frac{1}{1+\left(\frac{BZ}{\rho g_4^2}
+\frac{e^{2V}k^2\Phi}{2\kappa_4^2B\rho}\right)^2}\;,\label{heatcond}
\eea
 for the heat conductivities,\footnote{Note that in 
\cite{Alejo:2019utd} the notation $\kappa$ was used for the heat conductivity 
instead of $\bar\kappa$.} and
\bea
{\alpha_{xx}} & = &  s\rho
\frac{(2\kappa_4^2)\frac{g_4^4}{Z^2}e^{2V}
k^2\Phi}{(2\kappa_4^2)^2\frac{g_4^4}{Z^2}B^2\rho^2+(2\kappa_4^2 
B^2+\frac{g_4^2}{Z}e^{2V}k^2\Phi)^2}\cr
&=&\frac{s}{\rho 
B^2}\frac{\frac{e^{2V}k^2\Phi}{(2\kappa_4^2)}}{1+\left(\frac{BZ}{ \rho g_4^2}
+\frac{e^{2V}k^2\Phi}{2\kappa_4^2B\rho}\right)^2}\cr
{\alpha_{xy}} 
&=&2\kappa_4^2{sB}\frac{(2\kappa_4^2)\frac{g_4^4}{Z^2}
\rho^2+(2\kappa_4^2)B^2+\frac{g_4^2}{Z}e^{2V}
k^2\Phi}{(2\kappa_4^2)^2\frac{g_4^4}{Z^2}B^2\rho^2+(2\kappa_4^2 
B^2+\frac{g_4^4}{Z^2}e^{2V}k^2\Phi)^2}\cr
&=&\frac{s}{B}\frac{1+\frac{B^2Z^2}{\rho^2g_4^4}
+\frac{Z e^{2V} k^2\Phi}{g_4^22\kappa_4^2\rho^2}}{1+\left(\frac{BZ}{\rho g_4^2}
+\frac{e^{2V}k^2\Phi}{2\kappa_4^2B\rho}\right)^2}\;,\label{thermelcond}
\eea
for the thermoelectric conductivities, where we restored the dependence on 
the Newton's constant $\kappa_4$ and of $g_4^2$
with respect to \cite{Alejo:2019utd}. Here $e^{2V}, \Phi, Z$ and $W$ are defined at the horizon $r_H$ of the black hole, with $e^{2V}$ being the 
warp factor for the spatial part of the holographic boundary, $(x,y)$. 

Note that the effective squared coupling is $g_4^2/Z(r_H)$. Also note that at $\Phi(r_H)=0$, $\sigma_{xx}\rightarrow 0$, which means that 
\be
\label{weirdlimit}
\frac{\bar \kappa_{xx}/T}{\sigma_{xx}}\rightarrow \infty.
\ee

Let us make a few comments about this limit.

In a translationally invariant (clean) system without 
magnetic field both charge and heat conductivities are infinite, because the 
electrons can move without dissipation. This can be seen by taking first the 
limit $B\to 0$ and then $\Phi(r_H)\to 0$ in the above equations. For finite 
magnetic field, the gravity system should correspond to a strongly coupled 
theory of FQHE type on the holographic boundary. In such a system the FQHE bulk 
electron states are localized, unable to transport electric charge and, 
similarly, heat. Strictly speaking, for localization one needs disorder in the 
bulk that would create energy levels for the electrons to occupy. For this 
reason, in a perfectly clean limit, one expects that there is essentially no 
quantum Hall effect and the sequence of plateaux in the $\sigma_{xx}$ is 
replaced by a continuous dependence $\rho/B$. The sequence of peaks in 
$\sigma_{xx}$ in this limit can either be replaced by a finite value or vanish. 
In a real experiment one expects to observe finite  $\sigma_{xx}$: first, 
because the peaks are broad and second, because 
edge currents can contribute to $\sigma_{xx}$. In the holographic model, for 
$\Phi(r_H)=0$, $\sigma_{xx}=0$. 

In the meantime longitudinal heat conductivity does not vanish in the limit 
$\Phi(r_H)\to 0$. One notices that heat can also be transported by phonons, 
which do not carry charge. Phonons do not couple to the magnetic field, and we 
will see that for small temperature $\kappa_{xx}$ is ${\cal O}(T)$, independent from 
$\rho$ and $B$, consistent with the phonon interpretation. 

Another possible interpretation of~(\ref{weirdlimit}) in the clean limit 
$\Phi(r_H)\to 0$ is that the gravity dual describes an ensemble average over a 
disorder. Upon averaging, the translational invariance is restored in the 
system, but $\sigma_{xx}$ remains vanishing because it is so in every ensemble 
representative. Again, non-zero $\kappa_{xx}$ can be attributed to phonons. 
The ensemble average interpretation is consistent with the effective SYK-like 
description of the dyonic black hole that we discuss in the following part of 
the work.

In the general case, the \emph{modified longitudinal} Wiedemann-Franz law 
(the longitudinal Lorenz number $\bar L_{xx}$) reads
\be
\bar L_{xx}\  \equiv\ 
\frac{\bar \kappa_{xx}/T}{\sigma_{xx}}=\frac{s^2}{\rho^2}\frac{\frac{Z}{g_4^2}
+\frac{e^{2V}k^2\Phi}{(2\kappa_4^2)B^2}}{\frac{e^{2V}k^2\Phi}{(2\kappa_4^2)B^2}\left[
1+\frac{B^2Z^2}{\rho^2g_4^4}+\frac{Z e^{2V} 
k^2\Phi}{g_4^22\kappa_4^2\rho^2}\right]}\,.
\ee

{\em Note that we use $\bar \kappa_{xx}$ instead of the usual $\kappa_{xx}$}. 
The explanation will be given in the subsection after 
next.

If one takes the limit $B\to 0$, both the heat and the 
electric conductivities are infinite, but their ratio is finite for any 
$\Phi$. 
\be
\label{DirectLzeroB}
\bar L_{xx}\  = \ \frac{s^2}{
\rho^2+\frac{Z e^{2V} k^2\Phi}{g_4^22\kappa_4^2}}\,.
\ee

We will see in the subsection after next that, {\em with the assumption that 
we can use the dyonic black hole for the 
same calculation}, we can use the equations $s=c\a^2$, $\rho=\frac{c}{\pi}\a \mu$, $c\frac{g_4^2}{Z}=\pi$ and 
$\frac{\mu^2}{\a^2}-\frac{B^2}{\a^4}-3=\frac{4\pi T}{\a}$, and as $T\rightarrow 0$, cancel the dependence on $\a$, to obtain\footnote{$T\to 0$ properties of the conductivities in equations (\ref{chargecond})-(\ref{thermelcond}) were discussed in \cite{Melnikov:2012tb,Erdmenger:2016wyp,Melnikov:2017wfg}. In this work we provide new details about the connection with the conventional Wiedemann-Franz law.} 
\be
\frac{s^2}{\rho^2+\frac{B^2Z^2}{g_4^2}}=\frac{\pi^2}{3}+{\cal O}(T)\,.
\ee

It looks then that also in 
this case $\bar L_{xx}=\pi^2/3$ for $T\to 0$, at 
least for $\Phi\rightarrow 0$. That is possible if we take first $B\rightarrow 0$, {\em and then} $\Phi\rightarrow 0$. 
This might seem odd, given that $\Phi$ is a property of the model, whereas $B$ is an external field, but one can certainly take a 
very small but nonzero $\Phi$, yet $B=0$, in which case we obtain the desired 
result.

For the Hall components, the \emph{modified transverse} Lorenz number can be 
calculated for 
$\Phi(r_H)=0$. In this case, we obtain 
\bea
\sigma_{xy}&=&\frac{\rho}{B}-W(r_H)\,,\cr
\frac{\bar \kappa_{xy}}{T}&=&s^2\frac{\rho/B}{\rho^2+\frac{B^2Z(r_H)^2}{g_4^4}}\;,
\eea
so the modified transverse Lorenz number is 
\be
\bar L\equiv \frac{\kappa_{xy}/T}{\sigma_{xy}}=\frac{\rho/B}{\rho/B-4W}\frac{s^2}{\rho^2+\frac{B^2Z^2}{g_4^4}}.\label{WFwithW}
\ee

For $W(r_H)=0$ (no topological term, or see the discussion in the next subsection), and absorbing $Z(r_H)$ into $g_4^2$, we obtain 
\be
\bar L=\frac{s^2}{\rho^2+\frac{B^2}{g_4^4}}.\label{WFwithoutW}
\ee

We notice the similar structure of the Lorentz numbers~(\ref{DirectLzeroB}) 
and~(\ref{WFwithoutW}), in the longitudinal and transverse channels. Magnetic 
field in the transverse conductivity plays a role similar to disorder in the 
direct one. In the limit $B=\Phi(r_H)=0$ the Lorentz numbers are equal. 
However, at this moment, they do not yet look like the expected formula from 
the previous section.

What about the usual Lorenz numbers, in terms of  the thermal conductivity 
$\kappa$, instead of the heat conductivity $\bar \kappa$? The transition 
between the two is given by (\ref{kappabarkappa}). It only makes sense to cite 
the result at $W(r_h)=0$,\footnote{As will become clear below, the dependence 
on $W$ can also be recovered via a duality transformation.}
\begin{eqnarray}
\frac{\kappa_{xx}}{T} & = & \frac{s^2}{\rho}
\frac{\frac{\rho g_4^2}{Z}+\frac{k^2e^{2V}\Phi}{2\kappa_4^2\rho}}{B^2 + 
\left(\frac{\rho g_4^2}{Z}+\frac{k^2e^{2V}\Phi}{2\kappa_4^2\rho}\right)^2}\,, 
\\
\frac{\kappa_{xy}}{T} & = & - \frac{s^2}{\rho} 
 \frac{B}{B^2 + 
\left(\frac{\rho g_4^2}{Z}+\frac{k^2e^{2V}\Phi}{2\kappa_4^2\rho}\right)^2}\,,
\label{thermcond}
\end{eqnarray}

We note that if we take $\Phi=0$ first, we obtain the $\kappa_{xx}=\bar 
\kappa_{xx}$,  also $\kappa_{xy}\simeq 0$
now (at small $B$ and $W=0$). For $\kappa_{xx}$, the result for $\Phi = B = 0$ 
is finite, independent from the order of limits
\be
\frac{\kappa_{xx}}{T}\ \xrightarrow{B,\Phi\to 0} \ 
\frac{s^2}{\rho^2}\frac{Z}{g_4^2}\;,
\ee
and $L_{xx}$ either diverges, if one takes $\Phi\to 0$ first ($\sigma_{xx}=0$), 
or vanishes if instead $B\to 0$ first. For small $\Phi$ and $B=0$, 
\be
L_{xx} \ = \ 
\frac{s^2}{\rho^2}\frac{Z}{g_4^2}\frac{e^{2V}k^2\Phi}{(2\kappa_4^2)}\simeq 
\frac{\pi c}{3}\frac{e^{2V}k^2\Phi}{(2\kappa_4^2)}\,.
\ee

This small $\Phi$ result is also true for finite $W$. In the last step we 
substituted the values of $s$, $\rho$, $Z$ and $g_4$ for the dyonic black hole 
at $B\to 0$ and $T\to 0$ (as explained below).

For $\kappa_{xy}$, we can take $\Phi\rightarrow 0$ first, but keep $B$ and $W$ finite, and obtain 
\be
L=\frac{\kappa_{xy}}{T\sigma_{xy}}=-\frac{s^2}{\rho^2+B^2\frac{Z^2}{g_4^4}}\frac{\frac{\rho}{B}4W+\frac{Z^2}{g_4^4}}{(\rho/B-4W)^2}.
\ee

Using the fact that $Z/g_4^2=c/\pi$, we obtain 
\be
L=-\frac{1}{3}\frac{c^2+\frac{\pi^2 \rho}{B}4W}{(\rho/B-4W)^2}\simeq -\frac{c^2}{3\sigma_{xy}^2}\;,
\ee
where in the last equation we assumed that the second term is negligible with respect to the first.

In order to obtain and explain the dependence of the modified Lorenz number on 
$c$, note that, for an $AdS_4$ space, we have the holographic 
relation\footnote{In general, for $AdS_{d+1}$ we have 
\be
c=\frac{\pi^{d/2}R^{d-1}}{\Gamma(d/2)[l_P^{(d+1)}]^{d-1}}.
\ee
For an RG flow of the type 
\be
ds^2=e^{2A(r)}(-dt^2+d\vec{x}^2_{d-1})+dr^2\;,
\ee
we have 
\be
c=\frac{\pi^{d/2}}{\Gamma(d/2)[l_P^{(d+1)}A'(r)]^{d-1}}.
\ee}
\be
\frac{R^2}{2\kappa_4^2}=\frac{R^2}{16\pi G_N}=\frac{c}{4\pi}\;,
\ee
where $R$ is the $AdS_4$ radius, and in our gravitational ansatz, the left-hand side is 
$e^{2V}$ ($ds^2=...+e^{2V}(dx^2+dy^2)$). But on the other hand, the
entropy density of the field theory in $(2\kappa_4^2$ units), identified with the entropy 
of the black hole (Hawking formula) divided by the area in $(x,y)$, gives also 
(see also \cite{Alejo:2019utd})
\be
s=4\pi e^{2V(r_H)}=4\pi \frac{R^2}{2\kappa_4^2}=c.
\ee

That means that naively, the Lorenz number is proportional to $c^2$, if we think of 
$\rho, B$ and $g_4^2$ as constant parameters independent 
on the number of degrees of freedom (so the denominator in (\ref{WFwithoutW}) 
is constant). However, we have to remember that we reabsorbed 
$Z(r_H)$ into $g_4^2$, and the dependence on the horizon means a dependence 
on the theory, therefore on $c$. Indeed, in the subsection after the next one, 
we will find that this happens, and we get the correct expected Lorenz number. 

But before that, we need to understand how to act with dualities on the various 
transport coefficients.

\subsection{Action of $Sl(2,\mathbb{Z})$ duality and its interpretation}

In this subsection, we will consider the action of various dualities on transport in 2+1 dimensions.

In particular, we want to find the effect of electromagnetic S-duality and particle-vortex duality on the transport coefficients and 
on the Wiedemann-Franz law, and more generally, the effect of an $Sl(2,\mathbb{Z})$ duality group on the same. 

The action of particle-vortex duality in 2+1 dimensions on the conductivity coefficients (normal and Hall) is, as in \cite{Alejo:2019hnb} (see 
also \cite{Burgess:2000kj}, eq. 27, and 
\cite{Shapere:1988zv,Girvin:1984zz,Jain:1990sc,Jain:1992hi,Fradkin:1996xb} for 
some original discussion of the duality in quantum Hall physics),
\be
\sigma\equiv \sigma_{xy}+i\sigma_{xx}\rightarrow -\frac{1}{\sigma_{xy}+i\sigma_{xx}}=-\frac{1}{\sigma}\;,\label{PVlaw}
\ee
or 
\be
\sigma_{xx}\rightarrow \frac{\sigma_{xx}}{\sigma_{xx}^2+\sigma_{xy}^2}\;,\;\;
\sigma_{xy}\rightarrow -\frac{\sigma_{xy}}{\sigma_{xx}^2+\sigma_{xy}^2}.
\ee

On the other hand, the action of electromagnetic S-duality (understood as duality in 3+1 dimensions) on the 
same transport, is found from the equations derived in  \cite{Alejo:2019utd} for the holographic coefficients 
in the presence of $\rho,B$ and $W,Z$ in the dual, or in~\cite{Donos:2017mhp} in a similar calculation 
for holographic lattices. It corresponds in 2+1 dimensions to the same particle-vortex duality, 
and is denoted by an operation $S$.

The $S$ operation, together with the $T$ operation, combine to generate the group $Sl(2,\mathbb{Z})$, 
which in 2+1 dimensions acts on $\sigma$, as shown in \cite{Burgess:2000kj} as well. 
The action at the level of a Chern-Simons field theory was defined better in 
\cite{Witten:2003ya}.

Electromagnetic S-duality (so extended from fields to charges) in 3+1 dimensions is 
\be
(E,B)\rightarrow (B,-E)\;,\;\;\; (Q_e,Q_m)\rightarrow (Q_m, Q_e)\;,
\ee
which means that
\be
E/B\rightarrow -B/E\;.
\ee

Mapping holographically to 2+1 dimensions, $B$ remains $B$ (magnetic 
field in 3+1 dimensions is a source for magnetic field in 2+1 dimensions, 
as usual), while electric field $E$ in 3+1 dimensions sources charge density ($J^0$, time component of the current density). 
Then we also have 
the action of S-duality in the 2+1 dimensional case as
\be
\rho\rightarrow B\;,\;\; B\rightarrow -\rho\;,
\ee
so in particular we have the action
\be
\label{rhoBaction}
\frac{\rho}{B}\rightarrow -\frac{B}{\rho}.
\ee

 The classical couplings of the 3+1 dimensional gauge theory part of 
action~(\ref{MasterAction}) also transform under S-duality. As in the case of 
the conductivity, the most compact form is in terms of the complex combination
\be
\tau \ = \ -4W + i\frac{Z}{g_4^2}\;,
\ee
that transforms as $\tau\to - 1/\tau$, or
\be
\label{ZWtransform}
Z/g_4^2\rightarrow \frac{Z/g_4^2}{(Z/g_4^2)^2+(4W)^2}\;,\;\;\; 4W \rightarrow 
\frac{-4W}{(Z/g_4^2)^2+(4W)^2}\;,
\ee

It was noted in~\cite{Alejo:2019utd} that taking the limit $\rho=B=0$ in 
formulas~(\ref{chargecond}), while keeping $\Phi\neq 0$, results in 
$\sigma_{xx}=Z/g_4^2$ and $\sigma_{xx}=-4W$. So indeed, S-duality 
corresponds via holography to particle-vortex duality in 3 dimensions, acting as we
saw in (\ref{PVlaw}).

We observe that identification of conductivities with coupling constants 
of an effective theory resembles effective sigma model description of 
disordered conductors~\cite{Evers:2008zz}, in which the \emph{bare} coupling 
constant $g^2$ (here $g^2/Z$) is taken as the inverse classical Drude 
conductivity. Tuning the magnetic field, or charge density, one induces an RG 
flow, which has critical points, either at the plateaux, with $\sigma_{xx}=0$, 
or at the plateau transitions, $\sigma_{xx}\neq 0$.

From the point of view of the sigma model analysis, another particular point on 
our phase diagram, corresponding to $W=0$ and $\Phi=0$, is the strongly coupled 
plateau fixed point
\be
\sigma_{xx} = 0\;, \qquad \sigma_{xy} = \frac{\rho}{B}\;.
\ee

This fixed point is also invariant under S-duality~(\ref{PVlaw}) provided we
have the action~(\ref{rhoBaction}).

We can now analyze the phase diagram of the holographic model, as a function 
of 3 independent parameters and one RG scale. For this, it is 
convenient to redefine the variables in the conductivities, as these 
parameters and the RG scale identified as $\lambda$,
\be
\frac{\rho}{B} \to \hat{\sigma} \;, \qquad \frac{k^2e^{2V}\Phi}{2\kappa_4^2B^2} \ = \ 
\lambda\,, \qquad \zeta \ = \ \frac{Z}{g_4^2}\, \qquad w\ = \ -4W \,.
\ee

In terms of this parametrization, the electric conductivities read
\bea
\sigma_{xx}&=&\lambda 
\frac{\hat{\sigma}^2+\zeta^2+\zeta \lambda}{\hat{\sigma}^2 + \left({\zeta}
+{\lambda}\right)^2}\;,\cr 
\sigma_{xy} 
&=& w + \hat{\sigma}\frac{\hat{\sigma}^2+\zeta^2+2\zeta \lambda}{\hat{\sigma}^2 
+ \left({\zeta}
+{\lambda}\right)^2}\;. \label{RGeqs}
\eea

In these RG equations $\zeta$ and $w$ are the initial (bare) values of the 
conductivities at the scale $\lambda=\infty$. We will consider $\hat{\sigma}$ as 
a parameter of the theory. It is not hard to see that the RG flow bounds 
$w\leq {\sigma}_{xy}\leq w+\hat{\sigma}$. In the meantime $\sigma_{xx}$ is 
bounded from below by the arc
\be
\sigma_{xx}^2 + \left(\sigma_{xy} - \frac{\hat{\sigma}}{2} - w\right)^2 \ = \ 
\frac{\hat{\sigma}^2}{4}\,.
\ee

\begin{figure}
\includegraphics[width=0.45\linewidth]{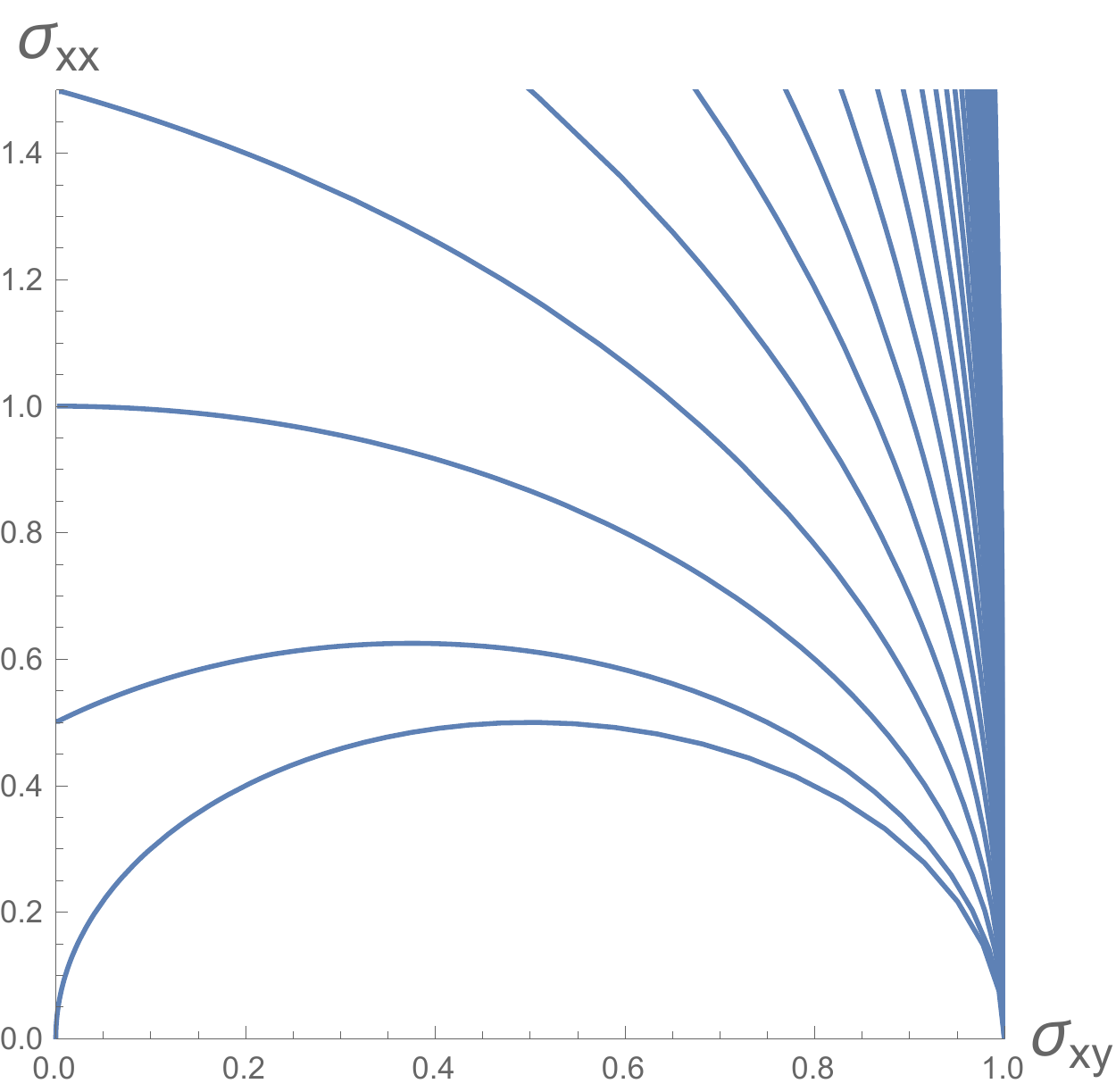}
{\hfill
\includegraphics[width=0.45\linewidth]{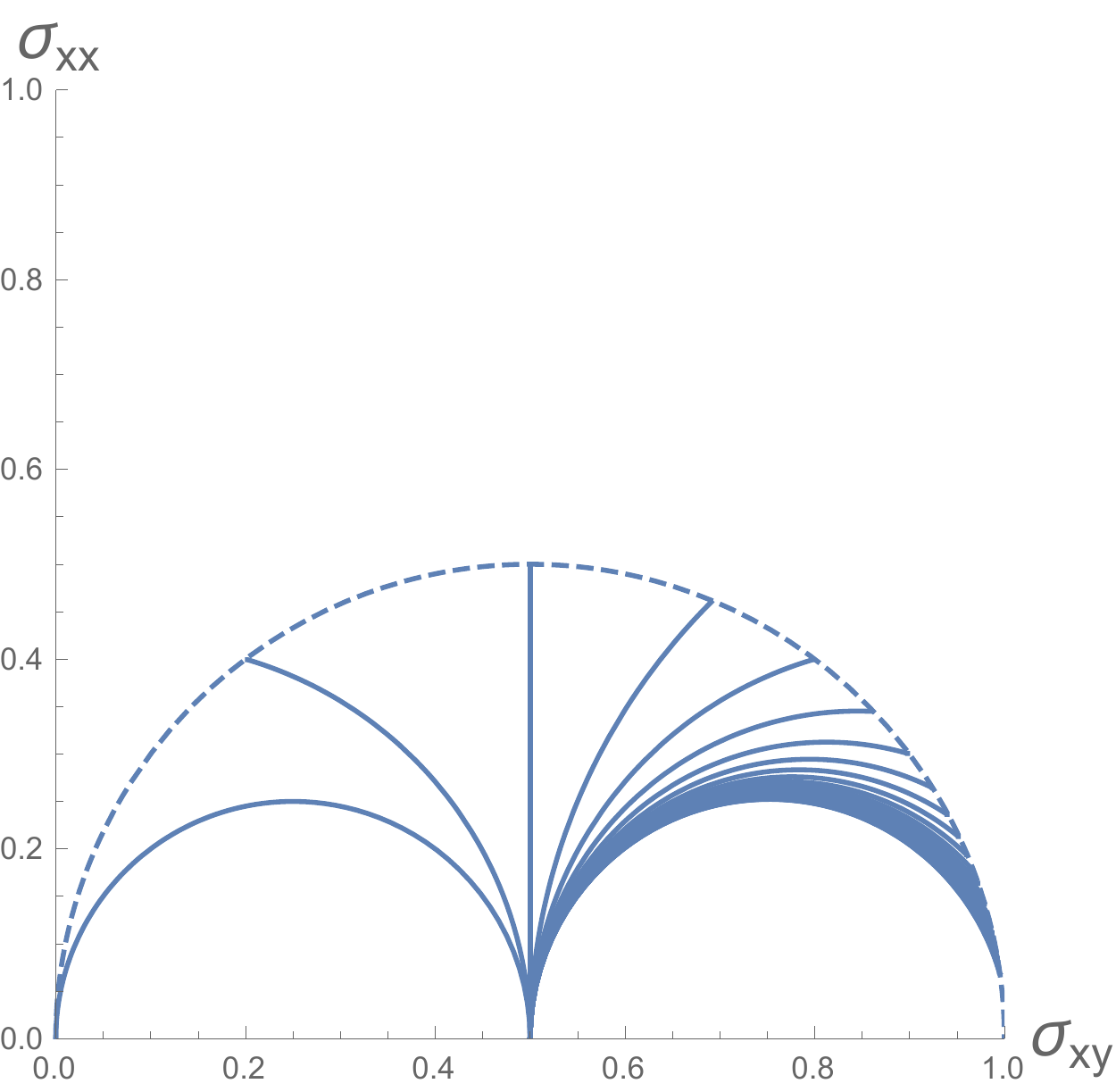}
}
 \caption{ (Left) Phase diagram of the flow in the space of conductivities 
fills the fundamental domain of the upper half plane in units of 
$\hat{\sigma}=\rho/B$. Different curves correspond to different bare values 
$\zeta$ of $\sigma_{xx}$. The bare value of $\sigma_{xy}$ is set to zero ($W=0$). 
(Right) The flow diagram of the theory upon the $STS$ transformation.}
 \label{fig:RGdiagram}
\end{figure}

If we rescale $\sigma_{xx}$ and $\sigma_{xy}$ to the units of $\hat{\sigma}$, 
the phase diagram is bounded by a region similar to the fundamental domain of 
the modular group $PSl(2,\mathbb{Z})$ as shown on figure~\ref{fig:RGdiagram}.

The phase diagram makes explicit the full duality structure of the theory. From 
\cite{Witten:2003ya} (page 2), note that for Abelian CS theories the T operation 
of $Sl(2,\mathbb{Z})$ acts as $\theta\rightarrow \theta+2\pi$, which means 
in our notation $W(r_h)\rightarrow W(r_h)+\hat{\sigma}/4$. Such shift simply 
translates the diagram on figure~\ref{fig:RGdiagram} to the next copy of the 
fundamental domain. Moreover, from eq. 4.6 in~\cite{Witten:2003ya}, from the 
current-current correlator in momentum space, 
from the $\delta_{ij}k^2-k_ik_j$ and the $\epsilon_{ijk}$ components, with 
coefficients $t$ and $w$, one creates $\tau=w+it$, on which S acts as 
$\tau\rightarrow -1/\tau$, and T acts as $w\rightarrow w+1$, as in the terms in 
the action above. Note that this is morally (if not rigorously)
related to the action on the conductivity, since by the Kubo relation the 
conductivity is the {\em retarded} current-current correlator 
in momentum space, divided by $i\omega$. 

Note that here, in the context of theories with holographic dual, 
the duality group is expected to be $Sl(2,\mathbb{Z})$ instead of 
$Sl(2,\mathbb{R})$, the reason being that in the holographic dual the 
corresponding electromagnetic black holes have integer charges $Q_e$ and $Q_m$.

Modular transformations map a point in the fundamental domain to an infinite 
set of dual theories (dual sets of values of the parameters). This action can 
also be understood as action on sets of initial conditions in the RG 
flow. At $W=0$ S-duality inverts the initial values of the RG flow (points on 
the vertical axis). Inside the fundamental domain the corresponding flows 
exchange accordingly. Indeed, equations~(\ref{chargecond}) transform 
as~(\ref{PVlaw}) provided~(\ref{rhoBaction}) and $Z/g_4^2\to g_4^2/Z$. T 
transformations, 
as said before, translate the flow horizontally between the neighboring 
domains.

For $W\neq 0$, transformations~(\ref{rhoBaction}) and~(\ref{ZWtransform}) no 
longer act on the conductivities as the S transformation of 
$Sl(2,\mathbb{Z})$. However, they still act properly on the 
initial values of the flow. An example of the flow generated by an $STS$ 
transformation of the initial conditions and of the flow 
equations~(\ref{RGeqs}) is shown on the right plot of 
figure~\ref{fig:RGdiagram}. It is not fully surprising that for a general copy 
of the domain, the S transformation is not defined in the same way as in the 
original domain at $W=0$. One can note that the difference with the $W=0$ 
transformation in this case, is that S transformation is not expected to act 
within the domain. The role of the latter transformation is played by 
$T^wST^{-w}$.

The next question is whether we can have an action of this $Sl(2,\mathbb{Z})$ on heat conductivity, 
and correspondingly on the Wiedemann-Franz law 
as well. 

{\em In the 2+1 dimensional field theory}, a self-duality of 
transport coefficients in $W=\Phi=0$ theory was noted in \cite{Hartnoll:2007ih} 
(there it also extended to the AC coefficients). The transport coefficients of 
their model were (note that we should actually replace $\omega\rightarrow 
\omega+i/\tau_{\rm imp}$,  were $\tau_{\rm imp}$ is the average time of 
scattering on impurities)
\bea
\sigma_{xx}=\sigma_Q\frac{\omega(\omega+i\gamma+i\omega_c^2/\gamma)}{[
(\omega+i\gamma)^2-\omega_c^2]}\,, &&
\sigma_{xy}=-\frac{\rho}{B}\frac{\gamma^2+\omega_c^2-2i\gamma\omega}{[
(\omega+i\gamma)^2-\omega_c^2]}\,,\cr
\rho_{xx}=\frac{1}{\sigma_Q}\frac{\omega(\omega+i\omega_c^2/\gamma+i\gamma)}{[
(\omega+i\gamma)^2-\omega_c^2]}\,,&&
\rho_{xy}=\frac{B}{\rho}\frac{
(\omega_c^2/\gamma)^2+\omega_c^2-2i(\omega_c^2/\gamma)\omega}{[
(\omega+i\gamma)^2-\omega_c^2]}\,,\cr 
\a_{xx}=\frac{\rho}{T}\frac{i\omega}{(\omega+i\gamma)^2-\omega_c^2}\,, &&
\a_{xy}=-\frac{s}{B}\frac{\gamma^2+\omega_c^2-i\gamma\omega}{
(\omega+i\gamma)^2-\omega_c^2}\,, \label{hydroresult1}\\
\theta_{xy}=-\frac{B}{T}\frac{i\omega}{(\omega+i\omega_c^2/\gamma)^2-\omega_c^2}
\,, &&
\theta_{xx}=\frac{s}{\rho}\frac{(\omega_c^2/\gamma)^2+\omega_c^2-i(\omega_c^2/\gamma)\omega}{(\omega+i\omega_c^2/\gamma)^2
-\omega_c^2}\,,\cr
\bar \kappa_{xx}=s\frac{i\omega-\gamma}{(\omega+i\gamma)^2-\omega_c^2}\,,&&
\bar\kappa_{xy}=-s\frac{\omega_c}{(\omega+i\gamma)^2-\omega_c^2}\,,\\
\kappa_{xx}=s\frac{i\omega-\omega_c^2/\gamma}{
(\omega+i\omega_c^2/\gamma)^2-\omega_c^2}\,,&&
\kappa_{xy}=s\frac{\omega_c}{(\omega+i\omega_c^2/\gamma)-\omega_c^2}\;,
\label{hydroresult3}
\eea
where $\sigma_Q$ is the electrical conductivity of the quantum critical system, $\omega_c$ has the interpretation as cyclotron frequency (and is indeed proportional to $eB$), and $\gamma$ is the damping frequency of the cyclotron modes.

 The claim of~\cite{Hartnoll:2007ih} is that the transformations
\be
\rho \leftrightarrow B\;, \qquad \sigma_Q\leftrightarrow 
\frac{1}{\sigma_Q}\;,\qquad  \gamma\leftrightarrow \frac{\omega_c^2}{\gamma}
\ee
imply the following duality on the transport coefficients that comes from exchanging $E_i\leftrightarrow \epsilon_{ij}J_j$ (sources with response fields) in the transport equations, 
\bea
\sigma_{xx}\rightarrow {\rho_{xx}} && \sigma_{xy}\rightarrow 
-{\rho_{xy}}\cr
\a_{xx}\rightarrow -\theta_{xy}=-(\hat \sigma^{-1}\hat \a)_{xy}&& \a_{xy}\rightarrow -\theta_{xx}\cr
\bar \kappa_{xx}\rightarrow \kappa_{xx}&& \bar \kappa_{xy}\rightarrow 
-\kappa_{xy}\;, 
\label{HKMSduality}
\eea
which includes an action on the heat conductivity. We remind the reader that the 
resistivities are defined as a matrix, inverse to the matrix of 
conductivities, $\hat{\rho}=\hat{\sigma}^{-1}$, which implies that
\be
\rho_{xx}\ = \ \frac{\sigma_{xx}}{\sigma_{xx}^2+\sigma_{xy}^2}\,, \qquad 
\rho_{xy}\ = \ -\frac{\sigma_{xy}}{\sigma_{xx}^2+\sigma_{xy}^2}\,,
\ee

However, that is not quite consistent due to some 
signs. First, we see that, in order to have a duality, in fact we need to change
\be
\sigma_Q\rightarrow \frac{1}{\sigma_Q}\;,\;\; \rho\rightarrow B\;,\;\; B\rightarrow -\rho\;,
\ee
(together with the same $\gamma\leftrightarrow \omega_c^2/\gamma$)
which is consistent with our description of the S-duality. Second, we note 
that, for $\sigma_{ij}=\sigma_{xx}\delta_{ij}+\sigma_{xy}\epsilon_{ij}$, 
$E_i\rightarrow \epsilon_{ij}J_j$ results in $\rho_{il}\leftrightarrow \epsilon_{ij}\sigma_{jk}\epsilon_{kl}=-\sigma_{il}$, 
or $\rho_{xx}\leftrightarrow 
-\sigma_{xx}$, $\rho_{xy}\leftrightarrow -\sigma_{xy}$. On the other hand, $E_i\leftrightarrow J_i$ results in $\sigma_{ij}\leftrightarrow
\rho_{ij}$, so we have in some sense half transformation from one duality, half from the other (but of course, that is not permitted). 
Finally, we note that in both cases, at $\alpha_{ij}=0$, we should have $\bar \kappa_{xy}\leftrightarrow \kappa_{xy}$, which is 
different than the above duality (\ref{HKMSduality}), where we have a minus sign.

One might think, why consider the formulas in \cite{Hartnoll:2007ih}, since they are for a model defined 
in 2+1 dimensions, and not  holographically? The reason is that the results for the transport coefficients  
match exactly!\footnote{ This match was partially observed in 
\cite{Hartnoll:2007ih}. Here we extend it to the case of non-trivial impurity scattering and, consequently, 
non-zero $\omega$.} First, the holographic formulas (\ref{chargecond}) and (\ref{heatcond}) (and even 
the thermoelectric ones, $\a_{xx}$ and $\a_{xy}$, in \cite{Alejo:2019utd})
are for DC transport, so at $\omega=0$, but we said that really $\omega\rightarrow \omega+i/\tau_{\rm imp}$, 
so for matching we should replace $\omega$ with $i/\tau_{\rm imp}$, but the breaking of translational invariance 
due to impurities is replaced in our case by $\Phi$, which does the same.\footnote{ In general the replacement 
is expected to be valid when $\omega\gg 1/\tau_{\rm imp}$. (We thank the referee for this point.) Although we seem 
to be in the opposite regime, the naive replacement still yields a consistent result.} Then (assuming that the $\Phi$ 
term should be identified with the $1/\tau_{\rm imp}$ term) indeed, the formulas for electric conductivities are 
the same, if (and only if) we identify 
\bea
&&\frac{1}{\tau_{\rm imp}\omega_c}=\frac{e^{2V}k^2\Phi}{2\kappa_4^2 B\rho}\;,\;\;\;
\frac{\gamma}{\omega_c}=\frac{BZ}{\rho g_4^2}\cr
&&\sigma_Q=\frac{Z}{g_4^2}.
\eea

We can also identify the formula for $\bar \kappa_{xy}$, provided we have
\be
\label{cyclotron}
\omega_c=\frac{\rho B}{sT}.
\ee

With these identifications (which already involved several nontrivial consistency checks), all other transport coefficients 
($\bar \kappa_{xx},\a_{xx},\a_{xy}$) are also matched. Then finally, we have 
the identification of parameters
\bea
\sigma_Q&=&\frac{Z}{g_4^2}=\zeta\cr
\omega_c&=&\frac{\rho B}{sT}=\frac{B^2}{sT}\hat \sigma\cr
\frac{1}{\tau_{\rm imp}}&=&\frac{e^{2V}k^2\Phi}{2\kappa_4^2}\frac{1}{sT}=\frac{B^2}{sT}\lambda\cr
\gamma&=&\frac{Z}{g_4^2}\frac{B^2}{sT}=\frac{B^2}{sT}\zeta\;,
\eea
where the second expressions are in terms of the reduced variables defined for the RG flow. We see that indeed, we had correctly identified the parameters, rescaled by $B^2/(sT)$, which is the factor that relates $\sigma_Q$ with $\gamma$. Second, $\lambda$, which is related to $\tau_{\rm imp}$ (the scale defined by impurities),  was correctly described as an RG scale, while also $\sigma_Q=\zeta$, the electric conductivity of the quantum critical system, was correctly described as value in the UV point ($\lambda\rightarrow\infty$). 

Then, given this match to the DC holographic formulas (\ref{chargecond}) and (\ref{heatcond}), we can extend them to the AC 
regime by replacing as in \cite{Hartnoll:2007ih}
\be
\lambda \frac{s}{\rho^2}=\frac{1}{\tau_{\rm imp}}\rightarrow \frac{1}{\tau_{\rm imp}}-i\omega.
\ee

In the above analysis we compared formulas~(\ref{chargecond})-(\ref{thermcond}), derived holographically in the approximation of weak electric and magnetic fields, with equations~(\ref{hydroresult1})-(\ref{hydroresult3}) derived hydrodynamically in the approximation of weak magnetic field. Consequently our values of $\omega_c$ and $\gamma$ match the hydrodynamic values of \cite{Hartnoll:2007ih} in the approximation $\mu\ll T$. On the other hand, we will see that equations~(\ref{chargecond})-(\ref{thermcond}) are also valid in the case of the dyonic black hole solution, for any $B$ and $\rho$, so a small disagreement between the results remains.

More recently, the hydrodynamic derivation of the conductivities was generalized to arbitrary values of $B$ in~\cite{Amoretti:2020mkp}.\footnote{We thank Daniel Brattan and Andrea Amoretti for updating us on this topic and sharing some of their unpublished results.} Here, to match equations (\ref{hydroresult1})-(\ref{hydroresult3}) we simply assumed that the frequency can be reinstated by a shift of $1/\tau_{\rm imp}$. The authors of~\cite{Amoretti:2020mkp} show that the correct $\omega$ dependence is captured by the higher order corrections in $B$, and that the hydrodynamic and holographic results match. It was also noted that conductivities, apart from $\sigma_Q$ also depend on another universal parameter $\sigma_H$ -- the incoherent Hall conductivity. This is consistent with the picture discussed in this work, since the ``flow" of the conductivities is two-dimensional, determined in terms of two initial values $\zeta$ and $w$.

As we saw in the example of Abelian CS of \cite{Witten:2003ya} there is 
a duality that acts on the parameters in the action, and a 
duality that acts in the resulting current-current correlators, or transport 
coefficients. In the first way of thinking $S$ and $T$ 
transformations act on parameters, with the functional form of the transport 
coefficients in terms of them constant {\em (passive duality)},  while in 
the second way the parameters remain unchanged but $S$ and $T$ act on the 
transport coefficients themselves {\em (active duality),} and the result is the 
same.

We can construct a heat analog of the complex conductivity,
\be
\frac{\bar\kappa}{T}\equiv 
\frac{\bar\kappa_{xy}+i\bar\kappa_{xx}}{T}=\frac{s^2}{B}\frac{\rho+i\left(\frac{
BZ}{
g_4^2}+\frac{e^{2V}k^2\Phi}{(2\kappa_4^2)B}\right)}{\rho^2+
\left(\frac{BZ}{g_4^2}+\frac{e^{2V}k^2\Phi}{(2\kappa_4^2)B}\right)^2}
=\frac{s^2}{B}\frac{1}{\rho-i\left(\frac{BZ}{g_4^2}+\frac{e^{2V}k^2\Phi}{
(2\kappa_4^2)B}\right)}.
\ee

Then the S-duality acts on it as 
\be
\frac{\bar\kappa}{T}\rightarrow 
-\frac{s^2}{\rho}\frac{1}{B+i\left(\rho\frac{g_4^2}{Z}+\frac{e^{2V}k^2\Phi}{
(2\kappa_4^2)\rho}\right)}=
i\frac{B}{\rho}\frac{Z}{g_4^2}\frac{s^2}{B}\frac{1}{\rho-i\frac{Z}{g_4^2}
B+\frac{Ze^{2V}k^2\Phi}{(2\kappa_4^2)g_4^2\rho}}.
\ee

So, while $\kappa/T$ looks nice even for nonzero $\Phi$, the transformation looks understandable as 
an active transformation only at $\Phi=0$,  or if 
\be
\Phi\rightarrow -i\frac{\rho}{B}\frac{g_4^2}{Z}\Phi
\ee

Then, 
\be
\bar\kappa \ \rightarrow \ i \frac{B}{\rho}\frac{Z}{g_4^2}\bar\kappa\,,
\ee
which amounts to
\be
\bar\kappa_{xx} \ \to \ 
-\frac{Z}{g_4^2}\frac{\bar\kappa_{xy}}{\sigma_{xy}}\,,\qquad 
\bar\kappa_{xy} \ \to \ \frac{Z}{g_4^2}\frac{\bar\kappa_{xx}}{\sigma_{xy}}\,.
\ee

This can be understood as a simple S-duality action on real ratios (rather than 
the complex combination $\kappa$ defined above)
\be
\frac{\bar\kappa_{xx}}{{\bar\kappa}_{xy}}\  \to \ 
-\frac{{\bar\kappa}_{xy}}{\bar\kappa_{xx}} \,.
\ee

In fact, from the point of view of self-duality~(\ref{HKMSduality}) 
observed in~\cite{Hartnoll:2007ih}, it is more natural to 
think of the duality transformation as action exchanging $\kappa$ with 
$\bar\kappa$ which is also an action on the real ratios, or as $2\times 2$ 
matrix action,
\be
\frac{\kappa_{xx}}{\bar{\kappa}_{xx}}\  \to \ 
\frac{\bar{\kappa}_{xx}}{\kappa_{xx}} \,, \qquad 
\frac{\kappa_{xy}}{\bar{\kappa}_{xy}}\  \to \ 
\frac{\bar{\kappa}_{xy}}{\kappa_{xy}}\,.
\ee

We remind the reader that $\bar\kappa$ and ${\kappa}$ are heat and thermal 
conductivities 
(the difference between them, expressed by equation~(\ref{heatvstherm}), is 
that the contribution to heat transport from electric fields must be subtracted 
from the latter). One can observe the simple duality 
$\bar\kappa_{ij}\leftrightarrow \kappa_{ij}$ directly in 
equations~(\ref{thermcond}), for 
any $\Phi$.

Finally we note again that for $W\neq 0$ the S-duality transformation 
makes sense only if $W(r_h)=n2\pi$. In this case, even including $\Phi$, the 
transformation above can be understood (instead of just S-duality) as a {\em 
passive transformation of $ST^n$}, where $T^n$ ($n$ T operations $W\rightarrow 
W+2\pi $) cancels out $W$, and $\Phi(r_h)$ is included as a parameter, as well 
as $e^{2V(r_h)}$, and are unchanged under the dualities.

It is also interesting to check the action 
of duality in the low temperature limit of the conductivities.  
\be
\bar\kappa_{xx} \ \to\  \bar\kappa_{xx} = \frac{\pi^2}{3}c T\,, \qquad 
\bar\kappa_{xy} \ = \ \frac{\pi^2}{3}c\frac{{g_4}^2}{Z}\sigma_{xy}T\ \to \ 
\kappa'_{xy} = \frac{\pi^2}{3}c\frac{{g_4'}^2}{Z'}\sigma'_{xy}T\,,
\ee
where
\be
\sigma'_{xy} = - \frac{B}{\rho} = - \frac{1}{\sigma_{xy}}\,,\qquad 
\frac{Z'}{{g_4'}^2} = \frac{g_4^2}{Z}\,.
\ee

We see that the direct heat conductivity is invariant under the 
action of S-duality, which agrees with  the above observation that 
$\bar\kappa_{xx}=\kappa_{xx}$ in the $\Phi=0$ limit. It also agrees with the 
non-electron (phonon) origin of the 
direct heat transport at low temperatures. On the other hand, the 
Leduc-Righi (transverse) conductivity is transformed in accordance with 
the particle-vortex picture. 

The action of S-duality, corresponding (modulo signs) to the exchange 
$E_i\leftrightarrow \epsilon_{ij} J_j$, namely 
$\sigma_{xy}\rightarrow 1/\sigma_{xy}$ for $\sigma_{xx}\simeq 0$,
$\bar \kappa_{xx}\rightarrow \kappa_{xx}$, $\bar \kappa_{xy}\rightarrow -\kappa_{xy}$, explains the exchange of the 
modified (transverse) Lorenz number $\bar L$ with the usual one $L$ only with a 
given dependence on $c$ for the case $cg_4^2=\pi Z$, 
namely if 
\be
\frac{\bar \kappa_{xy}}{T}=\frac{\pi c g_4^2}{3Z}\sigma_{xy}\Rightarrow \bar 
L=\frac{\bar\kappa_{xy}}{T\sigma_{xy}}=\frac{\pi^2}{3}\;,
\ee
as we obtained, being mapped by S-duality into
\be
-\frac{\kappa_{xy}}{T}=\frac{\pi 
c}{3}\frac{Z}{g_4^2}\frac{1}{\sigma_{xy}}\Rightarrow 
L=\frac{\kappa_{xy}}{T\sigma_{xy}}=
-\frac{c^2}{3\sigma^2_{xy}}\;,
\ee
where in the last equality we have used again $cg_4^2=\pi Z$.

\subsection{Wiedemann-Franz law from the dyonic black hole dual}
\label{sec:dyonic}

In this subsection, we will see that we indeed obtain the correct generalized 
Wiedemann-Franz law expected from the previous section, and calculate some 
interesting relations between thermodynamical quantities and conductivites. 

We will first summarize the calculations of the conductivities in 
\cite{Hartnoll:2007ai,Hartnoll:2007ih}, using their notation, which implicitly 
assumes $c g_4^2=\pi Z$, and then generalize to the case of the two parameters 
being independent. Assuming for simplicity $Z=1$, we see that in those 
works $1/g_4^2=1/(4\pi G_N)$ and radius $R=1$, which gives the central charge 
as $c=R^2/(4G_N)
=\pi/g_4^2$. We can see that by comparing their action, 
\be
S=\left(\frac{\sqrt{2}N^{3/2}}{6\pi}\right)\int d^4x\sqrt{-g} \left[-\frac{R}{4}+\frac{F_{\mu\nu}^2}{4}-\frac{3}{2}+...\right]\label{Stheirs}
\ee
with our (canonical) form of the action, 
\be
S=\int d^4x\sqrt{-g}\left[\frac{R}{16\pi G_N}-\frac{F_{\mu\nu}^2}{4 g_4^2}-\frac{6}{R^2}+...\right].
\ee

We also note that their $G_N$ is dimensionless and a function of only $N$.

The dyonic black hole in~\cite{Hartnoll:2007ai,Hartnoll:2007ih} is given by the 
metric
\be
ds^2 \ = \  \frac{\alpha^2}{z^2}\left(-f(z)dt^2 +dx^2 + dy^2\right) + \frac{1}{z^2}\frac{dz^2}{f(z)}\,,
\ee
where
\be
f(z) \ = \ 1+(h^2+q^2)z^4 - (1+h^2+q^2)z^3\,,
\ee
and the Maxwell field tensor is 
\be
F \ = \ h\alpha^2 dx\wedge dy + q\alpha dz\wedge dt\,.
\ee

Note that the horizon radius is $z=1$, but there is a parameter $\alpha$, 
which has a role similar to the inverse horizon radius. Identifying the chemical 
potential with the asymptotic value of the temporal component of the 
gauge potential and demanding that the component vanishes at the horizon gives
\be
\mu = - q\alpha\,.
\ee

The magnetic field is given by the flux of $F$ through the $xy$ plane, $B=h\alpha^2$.

The Bekenstein-Hawking temperature of the black hole is determined through the relation
\be
\label{DBHT}
\frac{4\pi T}{\alpha}\ = \ 3 - \frac{\mu^2}{\alpha^2}- \frac{B^2}{\alpha^4}\,.
\ee

Computing the Euclidean action on the black hole solution and equating it to the thermodynamic potential, one gets, 
after subtracting appropriate counterterms~\cite{Hartnoll:2007ih},
\be
\label{OmegaDBH}
\frac{\Omega}{V} \ = \  
c\frac{\alpha^3}{4\pi}\left(-1-\frac{\mu^2}{\alpha^2}+3\frac{B^2}{\alpha^4}
\right)\,.
\ee

From this expression one can compute the entropy density,
\be
\label{DBHs}
s=\frac{S}{V} \ = \  - \left.\frac{1}{V}\frac{\partial \Omega}{\partial T}\right|_{B,\mu} \ = \  
c\alpha^2\,,
\ee
where we obtain $T=T(\a, B,\mu)$ from (\ref{DBHT}) in order to calculate the 
partial derivative of $\Omega(T,\mu,B)$ at fixed $T$ and $\mu$, and the charge 
density
\be
\label{DBHrho}
\rho \ = \  - \frac{1}{V}\left.\frac{\partial \Omega}{\partial \mu}\right|_{B,T} \ = \  
\frac{c}{\pi}\alpha\mu\;,
\ee
where we do the same.

We are also interested in the second derivatives of the potential, giving the susceptibilities. 
One can similarly compute those derivatives, where after similarly using 
$T(\a,B,\mu)$ in the derivatives, we put $T\simeq 0$ 
in (\ref{DBHT}),  to find
\be
-\frac{1}{V}\left.\frac{\partial^2\Omega}{\partial\mu^2}\right|_{B,T} \ = \ 
\frac{6c\alpha_0^3/\pi}{6\alpha_0^2-\mu^2}  + O(T)\,,
\ee
and
\be
-\frac{T}{V}\left.\frac{\partial^2\Omega}{\partial T^2}\right|_{B,\mu} \ = \
\frac{4c\pi\alpha_0^3}{6\alpha_0^2-\mu^2}T + O(T^2)\,,
\ee
where we only evaluated the lowest order contribution at low temperature. 
The zero temperature value $\alpha_0$ is fixed from equation~(\ref{DBHT}).

The above two susceptibilities are the diagonal entries of  
the complete susceptibility matrix. We can also compute similarly the 
off-diagonal entries, to find the matrix
\be
\label{SusMatrix}
\chi_s \ = \ \left(
\begin{array}{cc}
\frac{6c\alpha_0^3/\pi}{6\alpha_0^2-\mu^2} &  
\frac{2c\mu\alpha_0^2}{6\alpha_0^2-\mu^2}\\
\frac{2c\mu\alpha_0^2}{6\alpha_0^2-\mu^2}T & 
\frac{4c\pi\alpha_0^3}{6\alpha_0^2-\mu^2}T
\end{array}
\right)\;.
\ee

According to the general response theory in the hydrodynamic 
limit~\cite{Kadanoff:1963}, the susceptibility matrix is related to the matrix 
of conductivities through relation~(\ref{Cond2Sus}), where $D$ is the matrix of 
diffusivities. Normally, in a field theory, like for instance the SYK case in 
\cite{Davison:2016ngz}, or the generic model used in 
\cite{Hartnoll:2007ai}, one calculates the dynamic susceptibilities $\chi(k,\omega)$ from 2-point functions of fluctuations in the 
model, and relating them with the static $\chi_s$ obtained from thermodynamics as above allows us to extract the diffusivity matrix $D$, 
and thus the conductivity matrix is found as $D\cdot \chi_s$. 

However, in the holographic system of the dyonic black holes in \cite{Hartnoll:2007ai,Hartnoll:2007ih} analyzed above, like in a 
different holographic model (with some translational invariance induced by some linear axions) in \cite{Davison:2016ngz}, the 
holographic conductivities $\sigma_{ab}, \a_{ab}, \bar\kappa_{ab}$ 
are directly found, from the holographic retarded Green's functions $G^R_{J_aJ_b}, G^R_{J_a\pi_b},G^R_{\pi_a\pi_b}$, respectively, 
by use of the Kubo formulas, or from calculations at the horizon (using the membrane paradigm). 
Then one conversely uses (\ref{Cond2Sus}) to find the diffusivity matrix $D$ from the conductivity matrix 
and the static susceptibility matrix $\chi_s$. This is what we want to do as well.

In subsection~\ref{sec:AdS/CMTandTransport} we considered the calculation of the 
conductivity matrix from a certain holographic model, which, like explained
for instance in Appendix I of \cite{Davison:2016ngz}, is {\em a priori} different than the dyonic black hole model, so for consistency
we should only use the formulas derived from it in \cite{Hartnoll:2007ih}. The dyonic black hole has no translational invariance 
breaking, so one obtains $\sigma_{xx}=0$, and from $G^R_{J_iJ_j}$ the classical Hall electric conductivity, $\sigma_{ij}=\frac{\rho}{B}
\epsilon_{ij}$, consistent with the results in 
subsection~\ref{sec:AdS/CMTandTransport} at $\Phi=0$. Yet for the other 
conductivities, the usual Kubo formulas do not apply anymore at $B\neq 0$ 
\cite{Hartnoll:2007ai}, but are modified according to \cite{Hartnoll:2007ih},
\bea
\a_{ij}&=&
\frac{1}{T}\frac{G^R_{J_iQ_j}}{i\omega} +\frac{M}{T}\epsilon_{ij}\cr
\frac{\kappa_{ij}}{T}&=&\frac{1}{T}\frac{G^R_{Q_iQ_j}}{i\omega}+\frac{2(M_E-\mu M)}{T}\epsilon_{ij}.\label{akaptheirs}
\eea

We obtain relations compatible with the ones obtained in \cite{Alejo:2019utd}, 
but in the $\Phi\rightarrow 0$ limit, with $\bar\kappa_{ij}$ 
from~(\ref{heatcond}) and $\a_{ij}$ from~(\ref{thermelcond})

Together, we have for $\Phi=0$
\be
\a_{ij}=\frac{s}{B}\epsilon_{ij}\;,\;\;\; \sigma_{ij}=\frac{\rho}{B}\epsilon_{ij}.\label{alpharho}
\ee

Also, from (\ref{heatcond}) for $\Phi=0$, we obtain 
\bea
\frac{\bar\kappa_{xx}}{T}&=&\frac{s^2}{\rho^2}\frac{\frac{Z}{g_4^2}}{1+\frac{B^2}{\rho^2}\frac{Z^2}{g_4^4}}\cr
\frac{\bar\kappa_{xy}}{T}&=& \frac{s^2}{\rho B}\frac{1}{1+\frac{B^2}{\rho^2}\frac{Z^2}{g_4^4}}\;,
\eea
the same result as derived from  (\ref{akaptheirs}). 

Given the fact that we used different ans\"{a}tze for a starting point, and the results were only valid at $\Phi=0$,
we should stop with the analysis of the dyonic black hole here. 

However, we will for the moment {\em assume} (which is by no means obvious) 
that we can consider the dyonic black hole as a limit of the analysis of 
subsection~\ref{sec:AdS/CMTandTransport}, and use the formulas obtained there, 
combining 
them with ones derived in this subsection for the dyonic black hole. 

In the holographic system of~\cite{Davison:2016ngz}, which is different 
than our case but the logic can be imported, as it relies only on general 
properties of transport and thermodynamics, it was found 
that the diffusivity matrix has the structure given by 
equation~(\ref{diffusivities}), with
\be
\label{Seebeck2}
2\pi{\cal E} \ \equiv \ \left(\frac{\partial S}{\partial \rho}\right)_T \ = \ -
\left(\frac{\partial \mu}{\partial T}\right)_\rho \ = \ 
\frac{4\pi^2}{3}\frac{c\alpha^2\rho}{c^2\alpha^4+B^2c^2+\pi^2\rho^2}.
\ee

The same structure of matrix $D$ was claimed to be reproduced by a 
multi-dimensional SYK model. Moreover, the equilibrium parameter ${\cal E}$ is 
directly related to the Seebeck coefficient,
\be
\label{Seebeck}
{\bar S} \ = \ \alpha\cdot\sigma^{-1} \ = \ 2\pi {\cal E}\,.
\ee
a nontrivial result, which however is claimed in  
\cite{Davison:2016ngz} 
to be equivalent to the form (\ref{diffusivities}) of the diffusivities matrix,
at least in the $T\to 0$ limit. We will now discuss the structure of the 
transport of the dyonic black hole,  with the stated caveat that we assume it arises as a limit of the 
general analysis in section~\ref{sec:AdS/CMTandTransport} 
in this limit.

The conductivities in this $cg_4^2/Z=\pi$ convention, from  
\cite{Hartnoll:2007ih}, (or calculated from equations~(\ref{chargecond}) 
and~(\ref{heatcond}) with $W=0$ and $\Phi=0$, if we replace $g_4^2/Z$ 
with $\pi/c$) are 
\begin{eqnarray}
\sigma_{xy} & = & \frac{\rho}{B}\,, \label{one}\\
\bar{\kappa}_{xy} & = &  
\sigma_{xy}\frac{\pi^2s^2T}{\pi^2\rho^2+c^2B^2}
\label{two}\,.
\end{eqnarray}

Taking the ratio of the conductivities and computing the lowest order 
contribution at $T\to 0$ by replacing $s$ in (\ref{DBHs}) 
and $\rho$ in (\ref{DBHrho}) into the above, and using (\ref{DBHT}) in the limit 
$T\rightarrow 0$,  one finds the modified Lorenz number (note that we use 
$\bar \kappa_{xy}$ instead of $\kappa_{xy}$ as we should have, as we 
already noted in section~\ref{sec:AdS/CMTandTransport})
\be
\label{BHLorenzxy}
\bar L=\frac{\bar{\kappa}_{xy}}{\sigma_{xy} T} \ = \ \frac{\pi^2 
s^2}{\pi^2\rho^2+c^2B^2} \ = \ \frac{\pi^2}{3} + O(T)\,.
\ee

We can also recall equation~(\ref{DirectLzeroB}) for the modified Lorenz number 
in the direct channel in the limit of zero magnetic field. For pure a 
system~($\Phi$=0) $L_{xx}$ is given by the same number: by a similar 
calculation, we find
\be
\label{BHLorenzxx}
\bar L_{xx} \xrightarrow{B,\Phi\to 0} \frac{s^2}{\rho^2} \ = \ \frac{\pi^2}{3} + 
O(T)\,.
\ee

The explanation of taking these modified Lorenz numbers $\bar L$ and $\bar L_{xx}$, 
and obtaining the free result is that, as we saw in the previous 
subsection, S-duality or particle-vortex duality exchanges (modulo some signs) $\kappa_{xy}$ with $\bar \kappa_{xy}$
and $\kappa_{xx}$ with $\bar \kappa_{xx}$, so 
we obtain the free result by using $\bar \kappa_{xy}$ and $\bar \kappa_{xx}$, which  
means basically that this strongly coupled calculation (as holographic 
dual to a perturbative gravity one) is S-dual to a free (or perturbative) 
system.

 The diffusivity matrix $D$ is computed by inverting the relation~(\ref{Cond2Sus}). 
Since we are dealing with a system with both electric and magnetic fields, we 
will consider both the conductivity and the diffusivity matrices to be complex,  
$\sigma=\sigma_{xy}+i\sigma_{xx}$ and $D=D_{xy}+iD_{xx}$, as described in 
(\ref{Dcomplex}). For the imaginary part, which corresponds to the longitudinal 
($xx$) conductivities, at any $T$,
\be
D^\parallel \ = \ \left( 
\begin{array}{cc}
 0 & 0 \\
 -2\pi {\cal E} D_2^\parallel & D_2^\parallel
\end{array}
\right), \qquad D_2^\parallel \ = \ 
\frac{3}{8}\frac{c^2\alpha^4+B^2c^2+\pi^2\rho^2}{
\alpha(B^2c^2+\pi^2\rho^2)}.\label{Dpar}
\ee

Note that this equation trivially implies $\sigma_{xx}=0$ and $\a_{xx}=0$, which 
is consistent with the above relations. 
But since $T\a_{xx}=0$, the lower left corner of the $D\chi_s$ matrix needs to vanish as well, and it does, since
we can check that 
\be
D_2^\parallel(-2\pi{\cal E}\chi_{11}+\chi_{21}) =0.
\ee

This seems nontrivial, but note that both $2\pi {\cal E}$ and $\chi_{ij}$ are calculated as thermodynamical coefficients. 
In fact, 
\be
\chi_{11}=-\left.\frac{\d^2\Omega}{V\d \mu^2}\right|_{B,T}=\left.\frac{\d\rho}{\d \mu}\right|_{B,T}\;,\;\;\;
\chi_{21}=-\left.\frac{\d }{\d\mu}\left.\left(\frac{\d \Omega}{V\d \mu}\right|_{B,\mu}\right)\right|_{B,T}=\left(\frac{\d
S}{\d \mu}\right)_{B,T}\;,
\ee
so the identity is actually 
\be
\left(\frac{\d S}{\d \mu}\right)_{B,T}=\left(\frac{\d S}{\d \rho}\right)_{B,T}\left(\frac{\d \rho}{\d \mu}\right)_{B,T}\;,
\ee
which is correct.

The form of $D^\parallel$ in (\ref{Dpar}) is the form of the 
matrix~(\ref{diffusivities}), found in 
\cite{Davison:2016ngz} to be obtained also in the SYK model,
for $D_1=0$. Indeed, 
in a clean, $\Phi=0$, gapped, $B\neq 0$, system one does not expect charge 
diffusion.

One also has a real component of the diffusion in this system, defined by the 
transverse ($xy$) transport,
\be
D^\perp \ = \ \left( 
\begin{array}{cc}
 \frac{\pi\rho}{2c \alpha B} & \frac{3c^2\alpha^4+3B^2c^2-\pi^2\rho^2}{8\pi 
c^2\alpha^3BT} \\
 \frac{\pi\alpha T(2B^2c^2+\pi^2\rho^2)}{2B(B^2c^2+\pi^2\rho^2)} & 
\frac{\pi\rho (3c^2\alpha^4-B^2c^2-\pi^2\rho^2)}{8c\alpha 
B(B^2c^2+\pi^2\rho^2)}
\end{array}
\right)\,.
\ee

It is not of the form~(\ref{diffusivities}).

Finally, from (\ref{alpharho}) in the $B\rightarrow 0$ limit, for the clean 
system at $\Phi=0$,
the Seebeck coefficient is simply~\cite{Melnikov:2017wfg}
\be
\bar{S}_{ab} \ = \ \frac{s}{\rho}\,\delta_{ab} \ = \ 
\frac{\pi}{\sqrt{3}}\delta_{ab} + O(T)\,.
\ee

The $T\to 0$ limit, for $B\rightarrow 0$ also, indeed agrees with the value of $2\pi{\cal E}$ in 
(\ref{Seebeck}), as conjectured in~\cite{Davison:2016ngz}. Yet it is 
puzzling that in this case we have 
$S=2\pi {\cal E}$, but no $D$ of the form (\ref{diffusivities}), when in \cite{Davison:2016ngz} they were 
claimed to be equivalent. We leave a full understanding of this fact for later.

We also add that 
the ${\cal O}(T)$ coefficient in $S_{ab}$ satisfies the so called Mott 
relation~\cite{Melnikov:2017wfg}.

Equations~(\ref{BHLorenzxy}) and~(\ref{BHLorenzxx}) precisely reproduce the 
canonical (weak coupling) Lorenz number from a holographic calculation. This 
was previously observed in \cite{Melnikov:2012tb} and further elaborated in 
\cite{Melnikov:2017wfg}.  

However, we note that, first, this result is 
obtained for a specific normalization of coupling constants, and, second, the 
Lorenz number was computed for the coefficient $\bar{\kappa}$, while the 
canonical one uses the thermal conductivity $\kappa$~(\ref{heatvstherm}),  
as noted above.

Since S-duality exchanges $\kappa_{xx}\to \bar{\kappa}_{xx}$ at low 
temperatures, the transverse 
conductivities yield the (unmodified) Lorenz number
\be
L \ = \ \frac{\kappa}{\sigma T} \ = \ \frac{c^2}{3 \sigma^2} + 
O(T)\,,
\ee
in this limit. We have already noted in the previous subsections that this is just S-dual to 
the free value for the modified Lorenz number. Let us discuss these issues in 
more detail.

The particular normalization used in~\cite{Hartnoll:2007ai,Hartnoll:2007ih} 
arises from a holographic model obtained by considering a consistent truncation 
of the low-energy limit of M theory on $AdS_4\times 
S^7$\cite{Herzog:2007ij,Itzhaki:1998dd}. This model flows to a 
superconformal fixed point in the IR, which is known as the ABJM 
model~\cite{Aharony:2008ug}. Hence it describes a universality class of $2+1$ 
dimensional systems. We will see however, that relation~(\ref{BHLorenzxy}) is 
not preserved by the duality. More generally, the Lorenz number will depend on 
$g_4^2$.

For independent $g_4$ and $c$, the above equations generalize as follows. 
From (\ref{Stheirs}) we see that in order to go to the canonical form with 
independent $c$ and $g_4^2$, 
we need to write $(A_\mu)^2=(\tilde A_\mu)^2\frac{Z}{g_4^2}
\frac{\pi}{c}$, so $B^2=\tilde B^2\frac{Z}{g_4^2}\frac{\pi}{c}$ and 
$\mu^2=\tilde \mu^2\frac{Z}{g_4^2}\frac{\pi}{c}$. 

Then the  blackening factor of the black 
metric and the temperature generalize to
\be
f(z) \ = \  1 - 
\left(1+\frac{Z\pi}{cg_4^2}\left(\frac{\mu^2}{\alpha^2}+\frac{B^2}{\alpha^4}
\right)\right)z^3  +    
\frac{Z\pi}{cg_4^2}\left(\frac{\mu^2}{\alpha^2} 
\frac{B^2}{\alpha^4}\right)z^4
\ee
and
\be
\frac{4\pi T}{\alpha} \ = \ 3- \frac{Z\pi}{c 
g_4^2}\left(\frac{\mu^2}{\alpha^2}+\frac{B^2}{\alpha^4} \right). \label{four}
\ee

The expression for the thermodynamics potential becomes
\be
\frac{\Omega}{V} \ = \  \frac{c\alpha^3}{4\pi}\left(-1 
+ \frac{Z\pi}{cg_4^2}\left(-\frac{\mu^2}{\alpha^2} + 
3\frac{B^2}{\alpha^4}\right)\right).
\ee

The expression for the entropy remains unmodified but a factor of $g_4^2$ 
is restored in the charge density,
\begin{eqnarray}
s & = & c\alpha^2\,, \label{DBHs2}\\
\rho & = & \frac{Z\mu\alpha}{g_4^2}\,.
\end{eqnarray}

For the second derivatives (the susceptibilities) the result is
\begin{eqnarray}
-\frac{1}{V}\frac{\partial^2\Omega}{\partial\mu^2} & = & \frac{3 \alpha Z\left(1+\frac{\pi Z}{cg_4^2}\left(  \frac{\mu ^2}{\alpha^2}
+\frac{ B^2}{\alpha^4}\right)\right)}{g_4^2 \left(3+\frac{\pi Z}{cg_4^2}\left(\frac{\mu^2}{\alpha^2}+3\frac{B^2}{\alpha^4}\right)\right)} 
\ \to \  \frac{6 \alpha^3 Z}{g_4^2 \left(6\alpha^2 - \frac{\pi 
Z\mu_0^2}{cg_4^2}\right)}\,, \label{BHsus1} \\
- \frac{T}{V} \frac{\partial^2\Omega}{\partial T^2} & = & \frac{2 \alpha ^2 c  \left(3- \frac{\pi Z}{ c g_4^2}\left(  
\frac{\mu ^2}{\alpha^2} +\frac{B^2}{\alpha^4}\right) 
\right)}{ \left(3+ \frac{\pi Z}{c g_4^2}\left(\frac{\mu ^2}{\alpha^2}+3\frac{B^2}{\alpha^4}\right)\right)} \ \to 
\ \frac{4c\pi\alpha^3  T}{ \left(6\alpha^2
- \frac{\pi Z\mu_0^2}{c g_4^2}\right)} \,, \label{BHsus2}\\
- \frac{T}{V} \frac{\partial^2\Omega}{\partial\mu\partial T} & = & \frac{4 \pi Z  \mu  T}{  g_4^2\left(3+ \frac{\pi Z}{c g_4^2}
\left(\frac{\mu ^2}{\alpha^2}
+3\frac{B^2}{\alpha^4}\right)\right)} \ \to \ \frac{2 \pi Z\alpha^2  \mu_0  T}{  
g_4^2\left(6\alpha^2- \frac{\pi Z\mu_0^2}{c g_4^2}\right)} \,. \label{BHsus3} 
\end{eqnarray}

We now consider the expressions (\ref{heatcond}) and~(\ref{chargecond}) for the 
conductivities, with $\Phi=W=0$, and obtain
\be
\bar{L} \ = \ \frac{s^2}{\rho^2+\frac{ZB^2}{g_4^4}} \ = \ 
\frac{c^2\alpha^4}{\frac{Z^2}{g_4^4}\left(\frac{\mu^2}{\alpha^2}+\frac{ZB^2}{
\alpha^4 }\right)\alpha^4} \ = \ 
\frac{(cg_4^2)^2}{Z^2(q^2+h^2)}\;,
\ee
but from (\ref{four}), we get that, as $T\rightarrow 0$, now 
$q^2+h^2=3cg_4^2/\pi Z$, so we obtain
\be
{\bar L} \ =\ \frac{\pi}{3}\frac{cg_4^2}{Z}\,.
\ee
which is not quite the expected  universal number. Under S-duality it is 
transformed as 
\be
{\bar L} \to \frac{\pi}{3}\frac{cZ}{g_4^2}\,,
\ee
so even in the special case of  the the top-down theory on M2 branes 
(ABJM), ${\bar L}$ is transformed into a ``strong coupling'' value with a 
structure less resembling equation~(\ref{KF}).

 However, we note that $g_4$, in the M2 brane normalization, is the 
coupling of the {\em gravity} theory, so the strong 
coupling regime is in fact a weak coupling regime of the dual {\em gauge} theory. We 
cannot, in general, expect gravity results to be valid for large $g_4$. 
However, we can expect the duality to hold and act on the conductivities as 
discussed in the previous section. Consequently, the duality exchanges the 
values of $\bar{\kappa}_{xy}$ and $\kappa_{xy}$, so in the weakly coupled gauge 
theory the value of the canonical Lorenz number $L$ is as it should be, 
provided by equation~(\ref{BHLorenzxy}).

In summary, the following picture emerges. In  the theory dual to the 
low-energy limit of M2 branes the weak coupling value of the electron Lorenz 
number is given by its canonical value $\pi^2/3$. When electron interactions 
are not negligible, the Lorenz number is modified.  In the strong coupling 
limit, $\kappa_{xy}$ is {\em inversely proportional to $\sigma_{xy}$, with the 
coefficient $c^2/3$.} In the meantime the dual vortex Lorenz number has the 
canonical value at strong coupling, where weakly interacting vortices replace 
electrons. Duality exchanges the values of the electron and vortex Lorenz 
numbers when going from weak to strong coupling. The M2 brane values do 
not appear universal and might be modified once a different gauge theory dual 
to the dyonic black hole is constructed.

Finally, we comment on
the  question we put before about the fact that at fixed $g_4^2$, 
since we have $s^2\propto c^2$, we 
expected $L\propto c^2$, because we understood $\rho$ and $B$ as being {\em applied} fields in 
the boundary theory and $g_4^2$ as a parameter. If we derive them from the 
gravity dual in terms of given $q$ and $h$, and take $T\rightarrow
0$, then it seems that indeed, we get the right answer, but why do that? We need to assume that 
\be
\rho^2+\frac{B^2}{g_4^4}=\frac{3c}{\pi}\a^4\frac{Z}{g_4^2}\propto c\;,
\ee
for fixed $\a$ and $g_4^2$.  The answer we believe is, as already hinted, 
that $Z\equiv Z(r_H)$ is the horizon value of a field, 
which corresponds to a variation of the  field theory on the boundary, allowing 
the dependence on $c$ of the denominator.

\section{Transport coefficients from one dimensional effective action for 
AdS/CMT holographic dual}
\label{sec:1dTranportCoeffs}

\subsection{One dimensional effective action for transport in holographic models and extensions of SYK}

In \cite{Davison:2016ngz,Sachdev:2019bjn}, an effective action has been 
proposed that includes charge in the SYK 
model, making it a model of complex fermions, 
and in the corresponding gravity dual theory: the 0+1 dimensional effective action involving the Schwarzian 
is related to black hole horizons in the case of the usual SYK 
mode \cite{Kitaev:2015,Maldacena:2016hyu}.

In \cite{Sachdev:2019bjn} it is explained how, from a 
{\em charged} black hole in $AdS_D$ background, with $D=d+2$, and in the 
near-horizon region, where we have
$\simeq AdS_2\times M_d$, by reduction on $M_d$, we can describe holographically the $AdS_2$ theory in terms of
a quantum theory in 0+1 dimensions, that also describes a complexification of the SYK model. 

The resulting 0+1 dimensional effective action is described 
in terms of the charge $Q$, temperature $T$, with parameters $K,\gamma$ and ${\cal E}$, that all describe the dynamics of the complex SYK model. We have 
the defining thermodynamics relations in the quantum mechanical theory
\bea
S(Q,T\rightarrow 0)&=&S_0(Q)+\gamma T+...\cr
\frac{dS_0(Q)}{dQ}&=& 2\pi {\cal E},\;\; T=0\;,
\eea
defining $\gamma$ and ${\cal E}$, while $K$ is the zero temperature compressibility,
\be
K=\frac{dQ}{d\mu}\;,\;\; T=0.
\ee

The imaginary time (0+1 dimensional) effective action, in the grand canonical ensemble and depending on two scalar fields $\phi$ and $f$, one of which, $f$, is a 
diffeomorphism, is 
\be
\label{EffAction}
I_{\rm eff}=-S_0(Q)+\frac{K}{2}\int_0^{1/T}d\tau (\d_\tau \phi+i(2\pi {\cal E}T)\d_\tau f)^2
-\frac{\gamma}{4\pi^2}\int_0^{1/T} d\tau \{\tan \pi Tf(\tau),\tau\}\;,
\ee
where $\{,\}$ is the Schwarzian, 
\be
\{ g(\tau),\tau\}=\frac{g'''}{g'}-\frac{3}{2}\left(\frac{g''}{g'}\right)^2.
\ee

\subsection{Calculating transport in the 0+1 dimensional action}
\label{sec:SYKcond}

In \cite{Davison:2016ngz}, the transport coefficients were calculated in 
a higher-dimensional theory made from multiple copies of the complex SYK model 
on 
a lattice, while it was noted that the transport in the 0+1 dimensional SYK model itself is trivial, 
since it gives constant coefficients (independent of the 
spatial momentum $k$). 

But we note that, if we consider the construction in the previous subsection, the transport coefficients 
of the 0+1 dimensional effective action are not actually 
trivial, in the sense that, while describing the complex SYK model, the effective action (\ref{EffAction}) 
is also an effective action for the near-horizon of the charged black hole in $AdS_{d+2}$, reduced over 
$M_d$. Yet the black hole itself is holographically 
dual to a $d+1$ dimensional field theory for some condensed matter system, as we considered 
in the first part of the paper. The black hole paradigm 
(implicitly used, since the calculations were done at the horizon of the black hole, by obtaining 
quantities that are independent of radius, so the 
calculation in the holographic UV region equals the calculation at the horizon) means that 
transport in the $d+1$ dimensional field theory is obtained 
from the near-horizon of the black hole. By reducing on the compact space $M_d$, 
we obtain the large distance (spatial momentum $k\rightarrow 0$) limit of the 
theory.  Then the 0+1 dimensional effective action (\ref{EffAction}) encodes the transport 
coefficients of the $d+1$ dimensional theory in the $k\rightarrow 0$ 
limit. To compare with the first part of the paper, note that the momentum squared $k^2$ 
appeared always multiplied by $\Phi$, so we should be able to compare 
with the equivalent $\Phi\rightarrow 0$ limit of the formulas.

It was also explained in~\cite{Davison:2016ngz} that $K$ and $\gamma$ are 
related via~(\ref{Cond2Sus}) to the matrix of charge and heat conductivities. 
Here we want to see that we can do calculate %these quantities 
 conductivities directly, by 
thinking of $I_{\rm eff}$ as the quantum effective action for transport in 
$d+1$ dimensions, reduced on the space dimensions (thus giving the zero spatial 
momentum limit of transport). 

If we think of $S$ as the response action of the quantum theory then
\be
\sigma^{ij}=\frac{\d j^i}{\d E^j}=\frac{\d(\delta  S/\delta A^i)}{\d \dot 
A_j}\;.
\label{sigma}
\ee

Indeed, $I_{\rm eff}$ in (\ref{EffAction}) is the quantum effective action 
reduced to 0+1 
dimensions, i.e., response action for a strongly coupled theory like the FQHE model in 2+1 dimensions. 

There are two fields in $I_{\rm eff}$, the diffeomorphism field $f$ that was present also in the 
Schwarzian action for the usual (real) SYK model, 
and the new field $\phi$, which according to~\cite{Sachdev:2019bjn} 
couples to charge, hence can be thought of as $A_0$, the zero component of the 
gauge field (reduced over the spatial 
dimensions). Another way of saying it is that $\phi$ is conjugate to charge, while $f$ is kept off-shell.
Then, we define 
\be
\sigma=\frac{\d \delta I_{\rm eff}}{\delta \phi \d \dot \phi}.
\ee

We note that this is not a trivial dimensional reduction statement derived from 
(\ref{sigma}), but rather must be understood as a dimensional continuation of 
(\ref{sigma}), but the unique one we can have in the 0+1 dimensional theory. 
Nevertheless, it is a dimensional continuation of a conductivity. On the other 
hand, in \cite{Davison:2016ngz}, $\sigma$ was calculated as a 2-point function, 
thus more like a susceptibility. Yet in 0+1 dimensions, because (\ref{chichisD}) 
becomes singular ($\chi(k,\omega)=0$, and independent of $D$), we cannot define 
a nontrivial $D$, thus because of (\ref{Cond2Sus}), the conductivities and 
susceptibilities are the same.  The same would be true in the 
relativistic or ``collisionless'' ($\omega\gg T$) regime of the higher 
dimensional theory, as explained in~\cite{Herzog:2007ij}.

Then we write the terms that have a  $\phi$ and a $\dot \phi$ in $I_{\rm eff}$. 
In particular, 
\be
\label{partint}
\frac{K}{2}\int_0^{1/T}d\tau (\d 
\phi+...)^2 = -\frac{K}{2}\int_0^{1/T}d\tau 
\phi\d^2\phi + \frac{K}{2}\phi \dot \phi\bigg|_0^{1/T} + \ldots\;,
\ee
where we can put all but one term on-shell, $\d^2\phi+...=0$, from which we 
find\footnote{Note that the factor of 2 is from taking the derivative of a 
square.} 
\be
\sigma(\tau)=\frac{\d \delta I_{\rm eff}/\delta A(\tau )}{\d \dot A(\tau)}= 
K[\delta(\tau-1/T)-\delta(\tau)].
\label{sigmatau}
\ee

Next, we consider the heat conductivity, and we proceed in a similar manner, by finding a certain term in the effective action. 
Now however, to find the heat conductivity, we must consider the second term in 
$I_{\rm eff}$. Indeed, heat transport is defined in a general dimension as
\be
Q_i=-\bar\kappa_{ij} \d_j T\quad \Rightarrow \quad \bar \kappa_{ij} =-\frac{\d 
Q_i}{\d(\d_j T)}\,,
\ee
where $Q^i= T^{\tau i} -\mu J^i$ is the energy flux density in the 
absence of charge currents, so that the dimensional continuation to 0+1 
dimensions gives, analogously to the conductivity case,
\be
\bar\kappa=-\frac{\d Q_\tau}{\d T'}.
\ee  

Considering that $E=-L=-dS/d\tau$, we obtain
\be
Q_i=- \frac{\d^2 E}{\d A_i\d \tau}\;,\;\;
E=-\frac{dS}{d\tau}\Rightarrow Q_i=\frac{\d^2 (dS/d\tau)}{\d A_i\d 
\tau} = \frac{\d^2 L}{\d A_i\d \tau}\;,
\ee
where $dA_i$ is an area vector for the direction of the energy current.

The dimensional continuation is easier now, since all we need to do is remove the area vector, 
\be
Q_\tau=\frac{\d L}{\d \tau}.
\ee

That means that, to obtain the heat 
current $Q$ defined in 0+1 dimensions, we look for the term with $\tau$ in 
$L$. Then, to obtain $\bar \kappa$, we look for the $(\d T) \tau$ or $T$  
terms in the Lagrangian $L$. Indeed, varying $Q_\tau=\d_\tau L$ with respect to 
$-\d_\tau T$ should give the dimensional continuation of $\bar\kappa_{ij}$, 
i.e., the (corrected) heat conductivity $\bar \kappa$. But since 
we want to calculate it at $T\rightarrow 0$, when 
$\bar \kappa\propto T$, in reality we look  for $T\dot T \tau$ and $T^2$ 
terms in $L$. 

But the diffeomorphism $f(\tau)$ includes the identity, so  $f(\tau)\rightarrow \tau +f(\tau)$ where with this replacement, $f(\tau)$ is 
now infinitesimal. 

Then, expanding the tangent at $T=0$, 
\be
\tan\left(\pi T (\tau+f(\tau))\right)\simeq \pi T(\tau+f(\tau)) +\frac{(\pi T)^3}{3}(\tau+f(\tau))^3+...\;,
\ee
and taking the Schwartzian and considering that $T'\tau\ll T$ and $f(\tau)\ll 
\tau$ (so that we can ignore the terms with $T''$ and with two or more $T'$s, 
and all terms with $f$), and doing the expansion, we find (after some algebra)
that
\be
L=-\frac{K}{2}(2\pi {\cal E} T)^2
-\gamma TT'\tau - \frac{\gamma}{2}T^2+...\;,
\ee
which means that the heat flux in the $\tau$ direction is
\be
Q_\tau=-\frac{dE}{d\tau}=\frac{\d L}{\d \tau}=-K(2\pi {\cal E})^2TT' - \gamma 
TT'+...\;,
\ee
where we have again neglected $T'\tau\ll T$. We obtain
\be
\label{bkappaSYK}
\bar \kappa = \frac{\d Q_\tau}{\d(-T')}=\gamma T + K(2\pi {\cal E})^2T\;,
\ee
where the last term is needed for the difference between $\bar\kappa$ and 
$\kappa$. 

As the final step, we compute the thermoelectric response.

The thermoelectric coefficient is defined in a general dimension as 
\be
\a_{ij}=-\frac{\d j^i}{\d\nabla^j T}=-\frac{\d \delta S/\delta A_i}{\d \nabla^j T}\;,
\ee
where as before, we understood $S$ as a response action, so that $j^i=\delta S/\delta A_i$. Since, as in the case of $\sigma$, 
$\delta S/\delta A_i$ is dimensionally continued to $\delta I_{\rm eff}/\delta \phi$ and, as for $\bar\kappa$, $\d \nabla^i T$ is 
dimensionally continued as $\d  \dot T$, we obtain that in 0+1 dimensions,
\be
\a=-\frac{\d \delta I_{\rm eff}}{\d \dot T \delta \phi}\;,
\ee
so we must look for the term with $\phi$ and $\dot T$.

Since $f(\tau)\rightarrow \tau+f(\tau)$, such that afterwards $f(\tau)$ is infinitesimal, and can be ignored, 
the cross term for the first integral in (\ref{EffAction}), is
\be
I_{\rm eff}=+iK\int_0^{1/T}d\tau\, \dot \phi(2\pi {\cal E}T)(1+\dot f)+...\;,
\ee
and partially integrating we obtain 
\be
I_{\rm eff}=+\left.iK\phi2\pi {\cal E}T\right|_0^{1/T}-iK2\pi{\cal E} \int d\tau\, \phi \dot T+...\;,
\ee
so that finally 
\be
\a=i2\pi {\cal E}K.
\ee

Since we expect $\a$ to be real, this means that ${\cal E}$ is in fact 
imaginary, so we simply reabsorb $i$ in the definition of ${\cal E}$.

Moreover, then we have 
\be
\kappa=\bar \kappa-T\frac{\a^2}{\sigma}=\bar \kappa - T(2\pi {\cal 
E})^2K= \gamma T\;,
\ee
where we have already assumed (as we will shortly see) that $\sigma=K$.

This is indeed the expected result, as also obtained in \cite{Davison:2016ngz} from susceptibilities, ending the calculation of the 
matrix of transport coefficients.

Until now, we have considered the transport coefficients calculated in coordinate (i.e., $\tau$ in 0+1 dimensions) space, 
but we know that in fact we need to calculate the 
transport coefficients as a function of frequency $\omega$, thus we must 
consider their 
Fourier transform. For $\kappa$ and $\a$, which are constant in $\tau$, this 
amounts to the DC component, which would be
a $\delta(\omega)$ in frequency space, if the integration region would be 
infinite. As it is, we have 
\bea
\kappa(\omega)&=& \gamma T\,\frac{e^{\frac{i\omega}{T}}-1}{i\omega}\rightarrow 
\gamma \cr
\a(\omega)&=&i2\pi {\cal E}K\frac{e^{\frac{i\omega}{T}}-1}{i\omega}\rightarrow i2\pi {\cal E}\frac{K}{T}\;,\label{oneoverT}
\eea
where on the right we have written the DC limit, of $\omega\rightarrow 0$. In this DC limit in frequency space, we see
the extra $1/T$ with respect to the coordinate space result, which is there just to make dimensions work, but otherwise we can 
drop it.
 
We are left to understand  $\sigma(\omega)$. 
Taking the Fourier transform of  (\ref{sigmatau}),\footnote{Note that in principle, the integration until $1/T$ assumes 
that time is periodic (functions of time are periodic with period $1/T$), but that is clearly 
not the case for $\sigma(\tau)$. Nevertheless, we continue with the calculation.}
\be
\sigma(\omega)=\int_0^{1/T}e^{i\omega\tau}\sigma(\tau).
\ee
we find
\be
\sigma(\omega)=K(e^{i\frac{\omega}{T}}-1)
\ee

Now, given that we want to take the limits $\omega\rightarrow 0$ and $T\rightarrow 0$, we 
assume that we can take it such that $\omega/T$=fixed and small.
Then we obtain 
\be
\sigma(\omega)\simeq Ki\frac{\omega}{T}.
\ee

If we think of the electric current as gauge/gravity dual to $\phi$ (since we 
want $\phi$ to represent a 1-dimensional gauge field $A_i$) then the 
conductivity (by the Kubo formula) is the retarded current-current correlation 
divided by $i\omega$, and we can think of $I_{\rm eff}$ as the gravity dual 
effective on-shell action, whose variation gives the current-current correlation 
function. This is consistent with the result for the DC values of $\kappa$ and 
$\a$ in (\ref{oneoverT}), where the $1/(i\omega)$ arose from the integration 
over $\tau$.

Then we note that the same factor of $1/T$ as in (\ref{oneoverT}), needed since 
otherwise dimensions don't work out in the transport coefficients, so we can 
drop it as well.

In this way, we get rid of the factor of $i\omega/T$, and we finally find
\be
\sigma(\omega\rightarrow 0, T\rightarrow 0) = K.
\ee

But $\gamma$ and $K$ are parameters, but not necessarily independent ones. In fact, we know that, 
if we take $I_{\rm eff}$ as the effective action
for a $d+1$ field theory at zero spatial momentum, they must be related by the 
Wiedemann-Franz law, 
\be
\label{1DL}
L\ = \ \lim_{T\rightarrow 
0}\frac{\kappa/T}{\sigma}=\frac{\gamma}{K}\,, \qquad \bar{L} 
\ = \ \lim_{T\rightarrow 0}\frac{\bar\kappa/T}{\sigma}=\frac{\gamma  + 
4\pi^2{\cal E}^2K}{K}\,.
\ee

As stressed before, here we are considering a different kind of $\kappa$ and 
$\sigma$. In the WF law of the holographic model of the previous sections, it 
is $\bar\kappa_{xy}$ and $\sigma_{xy}$. We also saw something similar for 
$\bar{\kappa}_{xx}$ and $\sigma_{xx}$. But in the calculation above, we rather 
had $\kappa_{00}$ and $\sigma_{00}$. An alternative derivation 
of~\cite{Davison:2016ngz}, which obtains $K$ and $\gamma$ from as correlators 
of the effective theory is reviewed in appendix~\ref{SYKApp}.

Then, how is it possible to have $\gamma$ and $K$ be parameters, 
and how is the WF law obtained? We have seen the WF law being obtained 
holographically directly from the $AdS_4$ dyonic black hole
gravity dual in the previous section. The point is that now $\gamma$ and $K$ are defined as parameters in the on-shell action, but in their 
relation to the $AdS_4$ theory, we have implicit the WF law.

\subsection{Comparing with Wiedemann-Franz law in $SYK_q$ and in 2+1 dimensions}

In the view that $I_{\rm eff}$ describes the SYK model, more precisely a 
SYK$_q$ generalization of the complex SYK model
(corresponding to $q=4$), with 
Hamiltonian \cite{Davison:2016ngz}
\be
H_0=\sum_{i_1<i_2<...<i_{q/2}; i_{q/2+1}<i_{q/2+2}<...<i_q=1}^N J_{i_1...i_q} \psi^\dagger_{i_1}\psi^\dagger_{i_2}...\psi^\dagger_{i_{q/2}}
\psi_{i_{q/2+1}}...\psi_{i_{q}}\;,
\ee
with $J_{i_1...i_q}$ random couplings with zero mean, giving a real Hamiltonian, 
and with constant modulus squared, 
\be
|J_{i_1...i_q}|^2=\frac{J^2(q/2)!^2}{N^{q-1}}\;,
\ee
it was obtained that the Wiedemann-Franz law is 
\be
\label{SYKL}
L=\lim_{T\rightarrow 0}\frac{\kappa/T}{\sigma}=\frac{\pi^2}{3}\frac{4}{q^2}.
\ee

Note that this result was obtained for a multidimensional extension of 
the SYK model, additionally with a specific choice of the intersite interaction. 
In terms of  the theory on M2 branes discussed in 
section~\ref{sec:dyonic}, the above result corresponds to the S-dual of the set 
of weakly-coupled gravity theories with couplings corresponding to 
${1}/({g_4^2})\to q^2/(4g_4^2)$. In other words we could replace $Z\to q^2/4$, 
with $q=2$ providing the canonical normalization. However, in the gravity 
description, $g_4$ and $Z$ are continuous parameters and the value of $L$ is 
not restricted to the discrete set. This is mirrored by the observation 
in~\cite{Davison:2016ngz} that the value~(\ref{SYKL}) is not universal and 
depends on the details of the intersite coupling of the multidimensional 
extension.

In this section we are discussing the $0+1$ dimensional model, for which $L$ 
and $\bar L$ are provided by~(\ref{1DL}). Consequently, we will compare 
the prediction of the $0+1$ dimensional SYK model with the results obtained from higher 
dimensional black holes reduced to $0+1$ dimensions. If $\gamma$ and $K$ are 
considered as continuous parameters in $I_{\rm eff}$, then $q$ must be 
(approximately) continuous too, which restricts it to be large, $q\rightarrow 
\infty$.

 The large $q$ limit of the SYK$_q$ model was analyzed in Appendix~C of 
\cite{Davison:2016ngz}: one assumes 
$J\rightarrow 0$, while keeping fixed 
\be
{\cal J}^2\equiv \frac{q^2 J^2}{2[2+2\cosh (\mu/T)]^{q/2-1}}\;,
\ee

In~\cite{Davison:2016ngz} the authors replace $\mu/T\to 2\pi {\cal E}$, but for 
us it will not be important since $\cal J$ will cancel from the final 
expression. It was further 
calculated (see eq. C.25 there)
\be
\gamma=\frac{2\pi^2(1-4Q^2)}{\cal J}\frac{1}{q^2}.
\ee

Moreover, if we first take $q\rightarrow \infty$ and then $T\rightarrow 0$, it was found that
\be
K=\frac{1-4Q^2}{4T}\;,
\ee
or, if we first take $T\rightarrow 0$ and then $q\rightarrow\infty$,
\be
K=\frac{q^2}{4{\cal J}}\;,
\ee
which means we obtain 
\be
\label{SYKL2}
L=\frac{\gamma}{K}=\frac{8\pi^2}{q^2}\frac{T}{\cal J}\,,\qquad \text{or} 
\qquad \frac{32\pi^2}{q^4}(1-4Q^2).
\ee

Since $\gamma$ and $K$ should be $T$-independent, it seems the latter limit 
($T\rightarrow 0$ first) is the needed one.  Consequently,
\be
\label{SYKbL}
\bar L \ = \ \frac{32\pi^2}{q^4}(1-4Q^2) + 4\pi^2{\cal E}^2 \ \simeq  \  
4\pi^2{\cal E}^2 \ = \  \left(\log\frac{1-2Q}{1+2Q} 
\right)^2 + O(1/q^2)\;,
\ee
where we have approximated in the case $Q$ fixed, $q\rightarrow \infty$.

In other words, $L$ and $\bar L$ have distinct limits and $q$ scaling for 
$q\to\infty$.

Now we use the alternative interpretation of $I_{\rm eff}$, as the effective 
action for the near-horizon limit of the $AdS_4$ black hole, holographically 
describing the 2+1 dimensional condensed matter field theory. 

 In the gravity dual described by a charged black hole, if one takes 
formulas (2.20) and (2.26) from~\cite{Sachdev:2019bjn}, for $d=2$, one has
\bea
K&=&\frac{s_2 R_h^{-1}[6R_h^2+\ell^2]}{3g_4^2}\cr
\gamma&=& \frac{4\pi^2 2s_2\ell^2R_h^3}{\kappa^2[6R_h^2+\ell^2]}\;, 
\label{gammaBH}
\eea
with $s_2=4\pi$ (but cancels out) and $\kappa^2=8\pi G$, and $c=\ell^2/(4G)$, 
where $\ell$ now is the radius of $AdS_4$ and $R_h$ the horizon radius, 
 so we obtain 
\be
\frac{\gamma}{K}=\frac{\pi 
cg_4^2}{3}\frac{\left(\frac{6R_h^2}{\ell^2}\right)^2}{\left(\frac{6R_h^2}{
\ell^2 } +1\right)^2  }.
\ee

Similarly,
\be
4\pi^2{\cal E}^2 \ = \ \frac{\pi 
cg_4^2}{3}\frac{\left(\frac{6R_h^2}{\ell^2}\right)\left(\frac{6R_h^2}{
\ell^2 } +2\right)}{\left(\frac{6R_h^2}{
\ell^2 } +1\right)^2 }\,.
\ee

But note that also  at zero temperature
\be
\frac{6R_h^2}{\ell^2}=\frac{\mu_0^2\kappa^2}{g_4^2}-2 \ = \ 
\sqrt{1+\frac{3g_4^2Q^2}{4\pi c}} - 1\;.
\label{Rhzero}
\ee

This formula makes sense since the coefficient of the Ricci tensor is 
$1/(2\kappa^2)$, and of the Maxwell term is $1/g_4^2$, while $A_{0}\propto 
\mu_0$ so the ratio $\mu_0^2\kappa^2/g_4^2$ is indeed dimensionless. 

In the holographic limit one considers $c\gg 1$ at fixed $Q$ (or 
$\mu_0^2\kappa^2/g_4^2\rightarrow 2$), in which case we have small black 
holes, $R_h\ll \ell$, so
\be
\label{gammaKratio}
L \ = \ \frac{\gamma}{K}\simeq \frac{\pi 
cg_4^2}{3}\left(\frac{3g_4^2Q^2}{4\pi c}\right)^2 \ = \ \frac{ 
3g_4^6Q^4}{16\pi c} \;,
\ee
and
\be
\label{gammaKratioBar}
\bar{L} \ = \ \frac{\gamma}{K} + 4\pi^2{\cal E}^2 \ \simeq 4\pi^2{\cal E}^2\simeq \ \frac{2\pi 
cg_4^2}{3} \left(\frac{3g_4^2Q^2}{4\pi c}\right) \ = \ \frac12\,g_4^4Q^2\,.
\ee

There are several ways one can compare the results with the SYK$_q$ predictions 
(\ref{SYKL2}) and (\ref{SYKbL}), while matching the large $q$ limit with the 
$R_h/\ell\rightarrow 0$ limit. 

If one chooses $c = {\cal O}(q^4)$ and $g_4={\cal O}(1)$ 
then one gets the correct $q$ scaling, but not quite the same dependence on the 
charge $Q$.

Note that the SYK$_q$ theory has two dimensionless parameters, 
$q\rightarrow \infty$ and $N$ and one dimensionful $J\rightarrow 0$ (${\cal J}$ 
and $\mu$ can be combined to make another dimensionless parameter, 
$Q\sim\mu_0q^2/{\cal J}$), compared to the two dimensionless and one 
dimensionful parameters in the gravity theory, $c$, $g_4$ and $\ell$ (the 
addional dimensionless parameter is $\mu_0\kappa$, or $R_h/\ell$ in the gravity 
solution, or $Q\sim \mu_0 \ell/g_4^2$). Note that in gravity, we 
first take $T\rightarrow0$, like in the $SYK_q$ theory, so we cannot use $T$ to 
make another dimensionless quantity. If we think of $\ell$ as $g_4$ 
independent and fixed, then $q\to \infty$ limit is analogous to the weak 
coupling limit $g_4\to 0$.  Also note that the extra parameter $q$ in 
the SYK model is needed to deal with the fact that we also have a {\em charge} in 
the gravity dual (see the parameter $Q$, or $\mu_0\kappa$), and correspondingly 
a charge in the condensed matter field theory. The usual SYK deals with a 
gravity dual without charge.

So a more natural way of treating equations~(\ref{gammaKratio}) 
and~(\ref{gammaKratioBar}) is to consider the limit 
$g_4^2c$ fixed, as in the example of the dyonic black hole. Then,
\be
L \ \sim \ \frac{(g_4^2c)^3Q^4}{c^4}\,, \qquad \bar{L} \ \sim \ 
\frac{(g_4^2c)^2Q^2}{c^2}\,.
\ee

The limit $g_4^2c\sim 1$ is expected to be the strong coupling regime, that 
cannot be compared with the weakly coupled field theory dual. However, this 
scaling can be achieved by choosing $c=O(q^2)$ and $g_4=O(1/q)$ for large $q$. 
In such a case $\bar L$ shows a behavior compatible with the 
ratio~(\ref{SYKL2}),
\be
\bar{L} \ \sim \ \frac{Q^2}{q^4}\,. 
\ee

 Let us repeat the analysis for the dyonic black hole of the previous section. 
First, consider as in  \cite{Davison:2016ngz} (eq. (2.41) there) that the 
susceptibilities of the SYK model, calculated as the second derivatives of the 
thermodynamic potential $\Omega$, and identified with the $\Omega$ calculated in 
the dual dyonic black hole, give the same result as the conductivities derived 
in the previous subsection, up to a constant, namely 
\be
\chi_s=\frac{1}{N}\begin{pmatrix} -\left(\frac{\d^2\Omega}{\d\mu^2}\right)_T & -\left(\frac{\d^2\Omega}{\d\mu\d T}\right)_\mu\\
-T\left(\frac{\d^2\Omega}{\d \mu \d T}\right)_\mu & -T\left(\frac{\d^2\Omega}{\d T^2}\right)_\mu\end{pmatrix}
=\frac{1}{N}\begin{pmatrix} K & 2\pi {\cal E} K \\ 2\pi {\cal E} K T & (\gamma+4\pi^2{\cal E}K)T\end{pmatrix}.
\ee

Indeed, this is what we calculated and argued for in the previous subsection. 
Then, the ratio of $\gamma$ and $K$ is obtained from the second 
derivatives of $\Omega$ for the dyonic black hole, found 
in equations~(\ref{BHsus1})-(\ref{BHsus3}), giving in the $T\rightarrow 0$ 
limit (and using also $\rho$ from (\ref{four}) and $B$ from 
(\ref{DBHs2}) in the limit $T\rightarrow 0$)
\be
\frac{\gamma}{K} \ = \ \frac{\pi 
cg_4^2/Z}{9}\left(6-\frac{\pi\mu_0^2}{c(g_4^2/Z)\alpha^2}\right) \ = \ \frac{\pi 
cg_4^2/Z}{3}\frac{(\rho^2g_4^4/Z^2+2B^2)}{(\rho^2g_4^4/Z^2+B^2)}.
\ee

Similarly,
\be
\frac{\gamma+4\pi^2{\cal E}^2K}{K} \ = \ \frac{2\pi cg_4^2/Z}{3}\,.
\ee

In particular, in the $B\to 0$ limit with $g_4,\rho$ fixed one gets
\be
L \ = \ \frac{\pi cg_4^2/Z}{3}\,, \qquad \bar{L} \ = \  \frac{2\pi 
cg_4^2/Z}{3}\,.
\ee

On the other hand, in any weak coupling scaling limit $g_4\to 0$, 
\be
L \ \simeq \ \bar{L} \ = \   \frac{2\pi cg_4^2/Z}{3}\,.
\ee

This is the property of equations~(\ref{SYKL2}) and $(\ref{SYKbL})$ at $Q=0$ 
(putting $Q=0$ before taking the limit $q\rightarrow\infty$). The values match 
if one identifies
\be
{cg_4^2/Z} \ = \ \frac{48\pi}{q^4}\,.
\ee

On the other hand, for the special point $cg_4^2/Z=\pi$ discussed in the case of 
the dyonic black hole, one obtains
\be
L(B=0) \ = \ \frac{\pi^2}{3}\,, \qquad L(B\neq 0) \ = \ \frac{2\pi^2}{3}\,.
\ee

\section{One dimensional effective action for transport with $Sl(2,\mathbb{Z})$ 
duality: electric/magnetic self-dual action and $\theta$ term}
\label{sec:1dEffAct}

In this section we generalize the effective action $I_{\rm eff}$ to one that is 
(manifestly) invariant under the $Sl(2,\mathbb{Z})$ duality of the first part 
of the paper. However, we usually have two approaches to deal with 
dualities: one is to write an action with both kinds of fields, like for 
instance in the case of the $Sl(2;\mathbb{Z})$ invariant type II B string theory 
action in 10 dimensions, and then consider field configurations with only one or 
the other of the fields, configurations that can be considered to be dual 
to each other (say, the fundamental, F1, string and the D-string, or D1-brane, 
in the above example). This would be a kind of active duality approach. 
The other is the "master action" approach, in which we write an action in terms 
of two different fields, and solving one or the other gives one description or 
another for {\em the same } physics. This is a kind of passive duality approach. 
We will have to decide which approach to take.

\subsection{Electric/magnetic self-dual action}

As a first step, we introduce magnetic charge coupling in $I_{\rm eff}$ 
(besides the electric charge coupling introduced by \cite{Davison:2016ngz}), 
in the process obtaining an effective action that is explicitly invariant under the S-duality operation. 

The introduction of magnetic charge coupling is in some sense trivial: we just 
need to add another field $\phi_2$  (that we can also think as a source of 
vorticity) with the same action and coupling term as $\phi_1$ (that coupled to 
electric charge), with a coupling of $\phi_1$ to $\phi_2$ which puts them on 
equal footing, and the conductivities derived from them must be inverse to each 
other, as dictated by active S-duality in the sense described above 
(remembering that the effective action is in 0+1 dimensions). 

On the other hand, we do have precedents for the passive duality in 
the terminology above: we know how to write a manifestly invariant action for a 
duality, for instance for T-duality on the worldsheet in 1+1 dimensions (in the 
Buscher sense), or particle-vortex duality in 2+1 
dimensions \cite{Murugan:2014sfa} or even Maxwell S-duality in 3+1 dimensions 
(see for instance \cite{Murugan:2016zal}). One writes a "master action" in 
terms of both the field strength of the original field, and the dual field, 
that acts as Lagrange multiplier for the Bianchi identity of the field 
strength. Then integrating instead the field strength of the original field, 
gives the dual action, in terms of what was previously a Lagrange multiplier. 
The coupling between the two fields is best understood in the case of the 
master action for worldsheet T-duality, in 1+1 dimensions, which 
is $\epsilon^{\mu\nu}b_\mu \d_\nu X$. In 0+1 dimensions, that would correspond 
to $T\phi_1\d_\tau \phi_2$, so this is one coupling term for the 
manifestly S-dual effective action we seek (the factor of $T$ is introduced 
for the dimensions to match with those of the other terms). Another 
possibility, of the same dimension as the other terms, 
is $\phi_1\d_\tau^2\phi_2$. Moreover,  the "electric field action" should 
have inverse coupling to the "magnetic field action" (as is always the case in 
these dualities via a master action for it). 

Now we have to choose between the active and passive type of S-duality 
invariant action. The first observation is that we need to distinguish between 
the high energy and low energy, or rather high/low frequency $\omega$ (since 
we are in 0+1 dimensions). We will start with the high energy one, since it is 
understood to be the fundamental one, in the RG flow description in 
2+1 dimensions, in the beginning of the paper. Note that the high energy 
action can also be viewed as the effective action in the limit, in which the 
$T\to 0$ is taken first.

We will take an intermediate point of view. Namely, we will mostly use the 
active mode, and write down an action containing both fields, but mostly 
consider situations with only one or the other. However, it could be that such 
a situation is an inconsistent truncation, and in that case, we need to 
integrate out one of the fields. In this case, however, we only expect the 
result to be correct for the conductivity $\sigma$, and not for $\a$ or 
$\bar\kappa$. For the latter, the full action must be used. 

Then, the high-energy action we propose is 
\bea
I_{\rm eff}&=&-S_0(Q)+\frac{K}{2}\int_0^{1/T}d\tau (\d_\tau \phi_1+i(2\pi {\cal 
E}T)\d_\tau f)^2 -\frac{\gamma}{4\pi^2}\int_0^{1/T} d\tau \{\tan \pi 
Tf(\tau),\tau\}\cr
&&+\frac{1}{2K}\int_0^{1/T}d\tau (\d_\tau \phi_2-iK(2\pi {\cal E}T)\d_\tau 
f)^2+\int_0^{1/T}d\tau \phi_1\d_\tau^2\phi_2.\label{leffhigh}
\eea

The action is invariant under the symmetry
\be
\phi_1\leftrightarrow \phi_2\;,\;\;K\rightarrow \frac{1}{K}\;,\;\; {\cal E}\rightarrow -{\cal E}K.
\ee

From the conductivity interpretation of the coefficients it makes sense to 
add an extra $K$ factor in the bracket of the second line and transform ${\cal 
E}$: this will ensure a relation like~(\ref{Seebeck}) for the coefficients and 
the duality between the thermoelectric and Seebeck coefficients, as 
in~(\ref{HKMSduality}).

The equations of motion of the above action with respect to $\phi_1,\phi_2$ are 
\bea
\d_\tau^2\phi_2&=& K\d_\tau\left[\d_\tau \phi_1+i(2\pi {\cal E} T)\d_\tau 
f\right]\cr
\d_\tau^2\phi_1&=&\frac{1}{K}\d_\tau\left[\d_\tau \phi_2-i(2\pi {\cal 
E}T)K\d_\tau f\right]\;,\label{eomhigh}
\eea
and they are solved by 
\be
\phi_2=0\;\;{\rm and}\;\; \d_\tau\left[\d_\tau \phi_1+ i(2\pi {\cal E}T)\d_\tau 
f\right]=0\label{reducedeom}
\ee
as well as 
\be
\phi_1=0\;\;{\rm and}\;\; \d_\tau\left[\d_\tau \phi_2-i(2\pi {\cal E}T)K\d_\tau 
f\right]=0\;,
\ee
meaning we can consistently put $\phi_1=0$ or $\phi_2=0$, as we wanted. 

On the other hand, for the full dynamics, we need to consider the action with 
both $\phi_1,\phi_2$. If we would use the equations of motion 
(\ref{eomhigh}) integrated once, with an integration constant $C$, and replace 
the resulting $\d_\tau\phi_2$ in the action, we would obtain an action 
where both $\phi_2$ and $\phi_1$ have vanished in the bulk,
\bea
I_{\rm eff}&=&-S_0(Q)+\frac{K}{2}\int d\tau 
\left[i(2\pi {\cal E}T)(\d_\tau 
f)\right]^2 -\frac{\gamma}{4\pi^2}\int_0^{1/T} d\tau \{\tan \pi 
Tf(\tau),\tau\}\cr
&&+K\phi_1\left[\d_\tau \phi_1+i(2\pi {\cal E}T)\d_\tau 
f\right]_0^{1/T}+\text{irrelevant terms}\;,
\eea
where one can also have some additional irrelevant terms due to the 
integration constant $C$. Thus, after integrating out $\phi_2$, also $\phi_1$ 
only enters in the boundary terms and has no dynamics. This is expected since 
the equations (\ref{eomhigh}) are degenerate. This is a consequence of the fact 
that the {\em bulk part} of the action only depends on the difference 
$\d_\tau\phi_1-\d_\tau\phi_2/K$. There is a gauge symmetry, which transforms 
$\phi_1\leftrightarrow\phi_2$, that is broken only by boundary terms. 

To see this, we can instead consider the following slight modification of 
(\ref{leffhigh}),
\bea
I_{\rm eff}&=&-S_0(Q)+\frac{K}{2}\int_0^{1/T}d\tau \left(\d_\tau 
\phi_1-\frac{1}{K}\d_\tau\phi_2+i(2\pi {\cal 
E}T)\d_\tau f\right)^2 \cr
&& -\frac{\gamma}{4\pi^2}\int_0^{1/T} d\tau \{\tan \pi 
Tf(\tau),\tau\}\,.
\label{Ieffhigh2}
\eea

This action has the same equation of motion as (\ref{leffhigh}) in the 
$\phi_1$ and $\phi_2$ sector, but is different in the coefficient of the 
$(\d_\tau f)^2$ term and in some boundary terms. At the same time, it makes the 
symmetry transparent and yields the expected transport coefficients.

 Indeed, in either the magnetic ($\phi_1=0$) or electric ($\phi_2=0$) gauges the 
high energy action (\ref{Ieffhigh2}) is just action~(\ref{EffAction}) 
for one of the fields with either $K$ and ${\cal E}$, or $1/K$ and $-{\cal 
E}K$. Following the procedures in the previous section one obtains the electric 
conductivity, in either electric, or magnetic regime,
\be
\sigma_{\rm electric}=K=\frac{1}{\sigma_{\rm magnetic}}\;.\label{sigmaelmagn}
\ee

The duality can again be interpreted either in the active (invert the 
conductivity), or in the passive (invert the parameter) way. Similarly, one can 
derive the thermoelectric coefficients considering in either $\phi_1=0$, or 
$\phi_2=0$ regime,
\be
\alpha_{\rm electric} \ = \ i2\pi{\cal E}K\,, \qquad \alpha_{\rm magnetic} \ 
\equiv S \ = \ -2i\pi{\cal E}\,.
\ee

The heat conductivity ${\bar\kappa}$ is the same in both pictures,
\be
\bar \kappa \ = \ \gamma T + K(2\pi {\cal E})^2T\;.
\ee

However, we wanted to have the possibility of an active view of the duality 
already reflected in the action, with {\em both } electric- and magnetic-type 
fields independently turned on. In this case, the action with full 
$\phi_1\leftrightarrow \phi_2$ gauge invariance is not good, yet the original 
one (\ref{leffhigh}), equivalent except for the boundary terms, is better. In 
order to get a good action, we replace the mixing term in (\ref{leffhigh}) with
\be
\int d\tau \left[\frac{1}{2}(\phi_1\d_\tau^2\phi_2+\phi_2\d_\tau^2\phi_1)\right]\;,\label{mixing}
\ee
differing with $\int d\tau \phi_1\d_\tau^2\phi_2$ only by the boundary terms $\left.\frac{1}{2}[\phi_1\d_\tau \phi_2-\phi_2\d_\tau
\phi_1\right|_0^{1/T}$. 
Assuming then that both $\phi_1$ and $\phi_2$ are turned on, and obey their equations of motion 
(\ref{eomhigh}), we use them to put on-shell the action, obtaining 
\bea
I_{\rm eff}&=&-S_0(Q)+\int_0^{1/T} d\tau\left[2\frac{K}{2}(2\pi {\cal E}T\d_\tau f)^2-\left(\phi_1-\frac{1}{K}\phi_2\right)K\d_\tau
(i2\pi {\cal E}T\d_\tau f)\right]\cr
&&-\frac{\gamma}{4\pi^2}\int_0^{1/T} d\tau \{\tan \pi 
Tf(\tau),\tau\}\cr
&&+\left[\frac{K}{2}\phi_1\d_\tau \phi_1+\frac{1}{2K}\phi_2\d_\tau \phi_2\right]_0^{1/T}+{\rm eqs. \;of\; m.}\label{highonshell}
\eea

However, we are mostly interested in the low energy limit 
$\omega\rightarrow 0$, since as we saw in the previous sections, we 
mostly take $T\rightarrow 0$, with $\omega/T\rightarrow 0$ (or fixed and small). 
In that case, the term $\phi_1\d_\tau^2\phi_2$ is negligible compared to the 
term $T\phi_1\d_\tau \phi_2$ (which was, on the other hand, negligible at high 
energies), that we argued is a natural one to appear in a S-duality invariant 
action. Moreover, this low energy action is an effective one, in which case 
the effect of the mixing term $\phi_1\d_\tau^2\phi_2$, potentially of the same 
order as the kinetic terms for $\phi_1$ and $\phi_2$, is taken into account. As 
such, the signs of the kinetic terms at low energy can be different than the 
ones at high energy. From the low energy effective action we expect the same 
symmetry that we observed in the previous sections, potentially 
acting differently than at high energies.

Therefore at low energies (frequencies) we propose the effective action
\bea
I_{\rm eff}&=&-S_0(Q)+\frac{K}{2}\int_0^{1/T}d\tau (\d_\tau \phi_1-i(2\pi {\cal E}T)\d_\tau f)^2
-\frac{\gamma}{4\pi^2}\int_0^{1/T} d\tau \{\tan \pi Tf(\tau),\tau\}\cr
&&-\frac{1}{2K}\int_0^{1/T}d\tau (\d_\tau \phi_2+iK(2\pi {\cal E}T)\d_\tau 
f)^2+T\int_0^{1/T}d\tau \phi_1\d_\tau\phi_2.
\label{Ieff2}
\eea

With this choice of signs the action is manifestly invariant under the 
transformations
\be
\label{Ieff2duality}
\phi_1\ \to\  \phi_2\,, \qquad \phi_2 \ \to \ - \phi_1\,, \qquad K \ \to \ - 
\frac{1}{K}\,, \qquad {\cal E} \ \to \ -{\cal E}K
\ee

In~(\ref{Ieff2}) one could equally consider 
transformations of $K$ and ${\cal E}$ without the sign change and keep the plus 
sign in front of the first term in the second line, like in the high energy 
case.

Variation with respect to $\phi_1$ and $\phi_2$ of the low energy effective 
action (\ref{Ieff2}) gives 
the equations
\begin{eqnarray}
T\d_\tau \phi_1 & = & \frac{1}{K}\d_\tau\left(\d_\tau\phi_2 + iK(2\pi{\cal 
E}T)\d_\tau f\right), \label{phi2eq}\\
T\d_\tau\phi_2 & = &  K\d_\tau\left(\d_\tau\phi_1 - i(2\pi{\cal E}T)\d_\tau 
f\right). \label{phi1eq}
\end{eqnarray}

 These equations are not degenerate, as compared to (\ref{eomhigh}), and 
one cannot realize the duality as a gauge symmetry. We can then consider 
integrating out one field or the other, as in the "master action" for the 
duality, describing the same physics from two different points of view. One can 
integrate out either $\phi_1$ and $\phi_2$ and obtain a purely 
electric or purely magnetic action, for example,  solving the $\phi_2$ 
equation of motion (\ref{phi2eq}),
\be
\d_\tau\phi_2 \ = \ K(T\phi_1-\mu - i(2\pi{\cal E}T)\d_\tau f)\,,
\ee
where $\mu$ is the integration constant. Then, 
\bea
I_{\rm eff}&=&-S_0(Q)+\frac{K}{2}\int_0^{1/T}d\tau (\d_\tau \phi_1-i(2\pi {\cal 
E}T)\d_\tau f)^2
-\frac{\gamma}{4\pi^2}\int_0^{1/T} d\tau \{\tan \pi Tf(\tau),\tau\}\cr
&&-\frac{K}{2}\int_0^{1/T}d\tau (T\phi_1-\mu)^2+ K\int_0^{1/T}d\tau 
\phi_1(T\phi_1-\mu - i(2\pi{\cal E}T)\d_\tau f).
\eea

If one sets $\mu=0$, then the equation of motion for $\phi_1$ obtained from 
this action, is the same as equation (\ref{phi1eq}) with $\phi_2$ on shell. 
Calculating the conductivies from this electric action gives the same results 
as the calculation in section~\ref{sec:SYKcond} with action~(\ref{EffAction}).

The magnetic action is obtained via the duality 
transformation~(\ref{Ieff2duality}) and the same conclusions apply for the 
magnetic conductivities.

It is not obvious what happens when we have both $\phi_1$ and $\phi_2$. One 
possibility, that we explore here, is to consider $\phi_1$ and $\phi_2$ as 
independent, and define a conductivity matrix (in electric/magnetic space; note 
that there is another matrix structure in terms of spatial directions, and yet 
another in terms of charge/heat conductivity), like a YM conductivity 
$\sigma_{ab}$. Such a Yang-Mills conductivity was considered in the 
introduction to \cite{Herzog:2007ij} (this paper extends Witten's paper about 
Abelian CS theories \cite{Witten:2003ya} to the nonabelian case, for $N$ 
M2-branes; note that this was before the construction of the ABJM model for 
M2-branes) and comes from the retarded 2-point function of the currents at 
$T=0$, 
\be
\left.\langle J_\mu^a(p)J_\nu^b(-p)\rangle_R\right|_{T=0}=\sqrt{p^2}\left(\eta_{\mu\nu}-\frac{p_\mu p_\nu}{p^2}\right) K_{ab}\;,
\ee
implying $K_{ab}=\sigma_{ab}$. Note that \cite{Herzog:2007ij} 
also says that S-duality acts as $K_{ab}\rightarrow (K^{-1})_{ab}$ (matrix inverse of the conductivity only in the YM adjoint index space).

Then, defining $A_i^a$ as the gauge field with 2-dimensional matrix indices in the electric/magnetic space, 
consider as a definition the conductivity matrix in the same indices, 
\be
\label{CondMat}
\sigma^{ab}=\frac{\d(\delta I_{\rm eff}/\delta A_j^b)}{\d A_i^a}.
\ee

Then, when using our generalized 0+1 dimensional effective action, the definition of the conductivity in this electric/magnetic space is 
the same, but for $A_j^a=\phi^a=(\phi_1,\phi_2)$. Then the electric conductivity in $\tau$ space is 
\be
\sigma^{11}=K[\delta(\tau-1/T)-\delta(\tau)]
\ee
and the magnetic conductivity in $\tau$ space is 
\be
\sigma^{22}=1/K[\delta(\tau-1/T)-\delta(\tau)]\;.
\ee

Note that there is no off-diagonal contribution to $\sigma^{ab}$ despite 
the presence of the $\phi_1\partial_\tau\phi_2$ term. In the computation, as in 
equation (\ref{partint}), one writes the action as a part proportional to 
equations of motion and a boundary term:
\be
T\int_0^{1/T}d\tau \phi_1\d_\tau\phi_2 \ = \ \frac{T}{2}\int_0^{1/T}d\tau 
\phi_1\d_\tau\phi_2 - \frac{T}{2}\int_0^{1/T}d\tau 
\phi_2\d_\tau\phi_1 + \frac{T}{2}\phi_1\phi_2\bigg|_{0}^{1/T}\,.
\ee

The first two terms will contribute to equations of motion and vanish on shell. 
The remaining boundary term will not contribute to the conductivities. The fact 
that $\sigma^{12}=\sigma^{21}=0$ makes it easier to correctly invert the 
conductivity.

However, as we noted previously, the $\tau$ space is not the correct way to 
calculate the conductivity, but rather the frequency space, where moreover we 
need to multiply by $T/(i\omega)$ for the electric-electric conductivity (and so 
also for the magnetic-magnetic one).
 
\be
\sigma=\begin{pmatrix} K +{\cal O}(\omega/T)& 0 \\ 
0 & \frac{1}{K}+{\cal O}(\omega/T)\end{pmatrix}.\label{Komega}
\ee

Then S-duality, $K\to -1/K$, acts indeed as $\sigma\rightarrow -\sigma^{-1}$.

For completeness we also consider the high energy case. Considering first 
the form of the action as (\ref{leffhigh}) with the mixing term replaced by 
(\ref{mixing}), from the on-shell form (\ref{highonshell}), we see already that 
again $\sigma^{12}=\sigma^{21}=0$, so we have in $\omega$ space (\ref{Komega}). 

Alternatively, we can consider the form of the action with explicit gauge 
symmetry in the bulk, and see the interpretation of the boundary terms we must 
add to it (to be equivalent with the above). The on shell boundary term of the 
action~(\ref{Ieffhigh2}) is
\bea
I_{\rm eff}&=& \frac{K}{2}\phi_1\left(\d_\tau 
\phi_1-\frac{1}{K}\d_\tau\phi_2\right)\bigg|_0^{1/T} - 
\frac{1}{2}\phi_2\left(\d_\tau 
\phi_1-\frac{1}{K}\d_\tau\phi_2\right)\bigg|_0^{1/T}\,.
\eea

We see that the boundary term contains mixing terms, which would 
lead to nonzero $\sigma^{12}$, which will spoil the transformation 
$\sigma\to\sigma^{-1}$ needed, so to fix that, and be consistent with the 
(\ref{leffhigh}) formulation, one needs to add to the Lagrangian the boundary 
term
\be
\label{Bterm}
\frac{1}{2}\left(\phi_2\d_\tau 
\phi_1 + \phi_1\d_\tau\phi_2\right)\bigg|_0^{1/T}\,,
\ee
which is the charge associated with the $\phi_1\leftrightarrow \phi_2$ 
symmetry. In accordance with the previous discussion this charge vanishes when 
either $\phi_1=0$ or $\phi_2=0$. We see then that the gauge symmetry is 
broken on the boundary by the term with the charge associated with it, meaning 
the $\phi_1$ and $\phi_2$ are true independent modes only on the 0-dimensional 
boundary of the 1-dimensional space (at the initial and final times).

\subsection{Adding a $\theta$ term: T operation}

Next we need to consider the T operation in $Sl(2,\mathbb{Z})$ duality, which 
should correspond to shifting a topological term in the effective action by a 
unit. Then, having the action of the S and T operations, we have true 
$Sl(2,\mathbb{Z})$ invariance. 

Since the T operation shifts, as we saw in the first part of the paper, the 
value of the theta term in 3+1 dimensional gravity action, or correspondingly 
the CS term in the 2+1 dimensional condensed matter field theory, we need to 
find an analog for this in the 0+1 dimensional $I_{\rm eff}$ for the field 
theory, that describes the $AdS_2\times M_2$ near-horizon gravity, reduced on 
$M_2$, and reduced on-shell to the 1-dimensional action for the boundary 
sources, i.e., the 0+1 dimensional response action. 

Since this is a response action for the 0+1 dimensional theory, giving the 
transport coefficients in 2+1 dimensions, it can only give the integer piece in 
the quantum Hall conductivity coefficient $\sigma_{xy}$. 

The 3+1 dimensional gravitational action is quadratic in $A_\mu$, so we have 
also a quadratic action in the on-shell 2+1 dimensional boundary action. When 
reducing $A_i$ to $\phi$ (for the zero spatial momentum part of the 
transport), we can think of the reduced field appearing in the theta term
as the phase (angle) in a complex field version of $\phi$. More exactly (see 
Appendix A) the part of the phase $\theta_{\rm vortex}$ corresponding to a 
vortex in 2+1 dimensions, with ansatz $\theta=\theta_{\rm vortex}=\theta_0+n\a 
$, where $\a$ is the polar angle on the spatial complex plane. 

Then the action we must add to $I_{\rm eff}$ must be 
\be
\label{ThetaTerm}
+W(r_h)\phi^2(r_h)\frac{2\pi}{T} \int_0^{1/T}d\tau (\d_\tau\theta)^2\;,
\ee
where $\tilde W(r_h)=W(r_h)\phi^2(r_h)$ is the object that is an integer (the 
3+1 dimensional gravitational action coupling $W(\phi)$ calculated at the 
horizon $r_h$, $\phi(r_h)$), and the quantity $\tilde W(r_h)/T\equiv \Theta$ is 
a constant giving a transport coefficient, that can be considered as a 
fourth constant in the effective action, besides $\gamma$, $K$ and ${\cal E}$. 
Indeed, then the equation of motion for $\theta$ is 
\be
\d_\tau^2\theta \ = \ 0 \qquad \Rightarrow \qquad  \theta\ =\ \theta_0+\a\tau 
\;,
\ee
and for consistency (so that we don't have a singularity) we impose 
periodicity (single-valuedness) of the phase $\theta$ in terms of the the 
periodic time $\tau$, giving $\a=mT$, with $m\in \mathbb{Z}$. 

Moreover, under the T operation of the duality, $W(r_h)\phi^2(r_h)\rightarrow W(r_h)\phi^2(r_h)+1$, which then leaves 
$ e^{iI_{\rm eff}}$ invariant, as we wanted. 

Finally, the Hall conductivity can be inferred as being the variation $\d_{\dot 
\phi}\delta I_{\rm eff}/(\delta \phi)$ for the vortex angle piece ($\theta_{\rm 
vortex}$) instead of the gauge field $A_i$ itself, giving  $\tilde W(r_h)$,
obtaining (as before, from the on-shell boundary terms, going to $\omega$ space, 
taking a limit, and dividing by $i\omega/T$) 
\be
\sigma_{xy}\simeq 2\pi \Theta \frac{i\omega}{T}\frac{T}{i\omega}= 2\pi \Theta.
\ee

That means that indeed, we have obtained the integer part of the quantum Hall 
conductivity, which was obtained from $W(r_h)$ of the 3+1 dimensional gravity 
dual at the horizon, as in the first part of the paper.

\section{Conclusions}
\label{sec:conclusions}

In this paper we have considered the Wiedemann-Franz laws and the action of 
$Sl(2,\mathbb{Z})$ in theories with an AdS/CMT holographic dual. We found that 
the holographic modified Lorenz number is $\bar L=cg_4^2k_B^2\pi/3$,  which
is electric-magnetic dual to the ordinary Hall (transverse) Lorentz number 
$L=ck_B^2\pi/(3g_4^2\sigma_{xy}^2)$.  In a theory that is self-dual under 
the action of electric-magnetic duality the modified number allows to access 
the weak coupling value from the strongly coupled gravity analysis. For the 
dual of the low-energy theory on M2 branes this value is $\pi^2/3$.

For the action of $Sl(2,\mathbb{Z})$, we first found that it gives constraints 
for the motion of the complex $\sigma=\sigma_{xy}+i\sigma_{xx}$ along an 
RG-like flow. Then we found the action  of the duality on $\sigma$ in the 
presence of nonzero $\rho$ and $B$, that  generalized the action at 
$\rho=B=0$ discussed in \cite{Alejo:2019utd}, and the action on the 
corresponding complex heat conductivity $\kappa=\kappa_{xy}+i\kappa_{xx}$, which 
is covariant, as in a passive version of duality. The holographic calculation 
matches with the generic expectation from field theory, both for the transport 
coefficients, and for the duality transformations. 

Moreover, we used a dyonic black hole calculation, assuming that it arises as a 
limit of the calculation in \cite{Alejo:2019utd}, to derive also the matrix of 
susceptibilities $\chi_s$ and of diffusivities $D$, and some relation between 
the various coefficients.  We tested the Kelvin relation~(\ref{Seebeck}) 
for the Seebeck coefficient conjectured 
in~\cite{Davison:2016ngz,peterson2010kelvin} and found that it holds in the 
$T\to 0$, $B\to 0$ limit, but without expected structure~(\ref{diffusivities}) 
for the diffusivities.

We have also found that we can calculate the transport coefficients and find the 
Wiedemann-Franz law from an effective 0+1 dimensional generalized Schwarzian 
action, valid for both the holographic duals, and a charged, generalized version 
of the SYK model. We then found a self-dual (under electric/magnetic duality, 
or S transformation) extension  of this 0+1 dimensional effective action, and 
how to add a theta term in order to describe the T transformation of 
$Sl(2,\mathbb{Z})$. This shows the power of the 0+1 dimensional effective 
action, to both describe the transport properties, and the $Sl(2,\mathbb{Z})$ 
invariance of the physics. In the $T=0$ effective theory, the latter can 
be realized as a gauge symmetry.

 Our study leaves a few open questions. It is interesting to know whether the weak coupling value of $L$ can be 
 obtained in other examples of holography, or whether the strong coupling value can be checked through a field 
 theory calculation. In particular, it would be interesting to study other candidates for self-dual theories. 
 We also found that holographic susceptibilities satisfy a relation similar to the Wiedemann-Franz law. 
 We are not aware whether a similar relation is known for susceptibilities in quantum matter. Finally, an 
 interesting question is whether the action of duality can be derived from the higher-dimensional SYK model 
 and whether it is compatible to the action discussed above. We leave the answers to these questions for a future research.

%%%%%%%%%%%%%%%%%%%%%%%%%%%%%%%%%%%%%%%%%%%%%%%%%%%%%%%%%%%%%%%%%%%%%%%%%%%%%%%%%%%%%%%%
\section*{Acknowledgements}
%%%%%%%%%%%%%%%%%%%%%%%%%%%%%%%%%%%%%%%%%%%%%%%%%%%%%%%%%%%%%%%%%%%%%%%%%%%%%%%%%%%%%%%%

We thank Luis Alejo, Prieslei Goulart and Ilya Gruzberg for useful discussions. We also thank Andrea Amoretti and Daniel Brattan for the correspondence regarding the magnetohydrodynamic derivation of the conductivities. The work of HN is supported in part by CNPq grant 301491/2019-4 and FAPESP grants 2019/21281-4 and 2019/13231-7. HN would also like to thank the 
ICTP-SAIFR for their support through FAPESP grant 2016/01343-7. The work of DM 
was supported by the CNPq grant 433935/2018-9. DM would also like to thank 
the FAPESP grant 2016/01343-7 for funding the visit to ICTP-SAIFR in August 
2019, when this work was initiated.

\appendix

\section{Review of particle-vortex duality in 2+1 dimensional field theory}
\label{sec:app1}

This section reviews the particle-vortex duality in the formulation started in \cite{Burgess:2000kj}, made more precise in \cite{Murugan:2014sfa}
(as a transformation of the path integral of a field theory), and used for transport in \cite{Alejo:2019hnb,Alejo:2019utd}.

Particle-vortex duality can be written as a transformation at the level of the path integral in 2+1 dimensions, 
possible since the action is made to be 
quadratic in fields. In it, one writes a master action for the duality in terms of two fields, and when eliminating one, we get the original 
action, when eliminating the other, we get the dual action. 

The action that needs dualizing is for a complex field $\Phi$ minimally coupled to a Maxwell field $a_\mu$ as
\be
S=\int d^3x \left[-\frac{1}{2}|D_\mu \Phi|^2-V(|\Phi|)-\frac{1}{4}F_{\mu\nu}^2\right]\;,
\ee
where $F_{\mu\nu}=\d_\mu a_\nu -\d_\nu a_\mu$. Writing $\Phi=\Phi_0 e^{i\theta}$, the path integral is over $\Phi_0, \theta$ and $a_\mu$. 

In the path integral, we can have vortex configurations, with ansatz $\Phi_0(r) e^{i\theta(\a)}$, $\theta(\a)=N\a$, where
$(r,\a)$ are the polar coordinates in 2 spatial dimensions, in which case $\theta$ is singular, so one splits
\be
\theta=\theta_{\rm smooth}+\theta_{\rm vortex}\;,
\ee
with $\epsilon^{ab}\d_a \d_\b \theta_{\rm smooth}=0$, but $\epsilon^{ab}\d_a \d_b \theta_{\rm vortex}\neq 0$.

We replace $\d_\mu\theta$ with an independent variable $\lambda_\mu$, but with the constraint that 
$\epsilon^{\mu\nu\rho}\d_\nu \lambda_\rho=0$, imposed by Lagrange multipliers $b_\mu$, leading to the master action
\bea
S_{\rm master}&=&\int d^3x \left[-\frac{1}{2}(\d_\mu\Phi_0)^2-\frac{1}{2}(\lambda_{\rm \mu,smooth}+\lambda_{\rm 
\mu, vortex}+ea_\mu)^2\Phi_0^2+\epsilon^{\mu\nu\rho}b_\mu \d_\nu \lambda_{\rm \rho, smooth}\right.\cr
&&\left. -V(\Phi_0)-\frac{1}{4}F_{\mu\nu}^2 \right].
\eea

In the path integral, the integral over $\theta$ has been replaced by an integral over $\lambda_\mu$ and $b_\mu$. 

Integrating over $b_\mu$ leads to $\lambda_\mu=\d_\mu$ of something, which I can call $\theta$, leading to the 
original action, in the explicit form with kinetic term 
\be
-\frac{1}{2}\int d^3x \Phi_0^2(\d_\mu \theta_{\rm smooth}+\d_\mu \theta_{\rm vortex}+ea_\mu)^2.\label{kinori}
\ee

If instead I path integrate over $\lambda_\mu$, which is equivalent (since we have a quadratic action) to using the classical solution
\be
(\lambda_\mu+ea_\mu)\Phi_0^2=e\epsilon^{\mu\nu\rho}\d_\nu b_\rho\;,
\ee
and replacing back in the master action, we obtain the dual action, 
\be
S_{\rm dual}=\int d^3x \left[-\frac{(f_{\mu\nu}^b)^2}{4\Phi_0^2}-\frac{1}{2}(\d_\mu \Phi_0)^2-e \epsilon^{\mu\nu
\rho}b_\mu \d_\nu a_\rho-\frac{2\pi}{e}b_\mu j^\mu _{\rm vortex}-V(\Phi_0)-\frac{1}{4}F_{\mu\nu}^2\right]\;.
\ee

While in the original action we have the field $\theta$, with coupling $1/\Phi_0^2$, as seen in (\ref{kinori}), in the dual action 
above we have the field $b_\mu$, with coupling $\Phi_0^2$ (assuming $\d_\mu\Phi_0=0$). The change in fields is
\be
\d_\mu \theta+ea_\mu=\frac{1}{\Phi_0^2}\epsilon^{\mu\nu\rho}\d_\nu b_\rho\;,
\ee
which is nothing but Poincar\'{e} duality in 3 dimensions. 

Moreover, the duality exchanges the electric current, 
\be
j_\mu=e\Phi_0^2\d_\mu\theta
\ee
with the vortex current
\be
j^\mu_{\rm vortex}=\frac{e}{2\pi}\epsilon^{\mu\nu\rho}\d_\nu \d_\rho \theta=\frac{1}{2\pi \Phi_0^2}\d_\nu j_\rho.
\ee

More generally, we can consider an arbitrary function $K(\phi)$ multiplying a kinetic term and thus acting as 1 over the coupling squared, 
for a master action
\be
S_{\rm master}=\int d^{2+1}x\left[-\frac{1}{2}(\partial_i\phi)^2-V(\phi)-\frac{1}{4}K(\phi)f_{ij}f^{ij}
+\frac{1}{2}\epsilon^{ijk}\chi \d_i f_{jk}\right].
\ee

Eliminating the Lagrange multiplier $\chi$ leads to the action for a scalar coupled to a gauge field,
\begin{equation}
\mathcal{L}_{\phi a_i}=-\frac{1}{2}(\partial_i\phi)^2-V(\phi)-\frac{1}{4}K(\phi)f_{ij}f^{ij}, \phantom{....}i,j=0,1,2,
\end{equation}
while eliminating $f_{ij}=\d_i a_j-\d_j a_i$ via 
\be
f^{ij}=-\frac{1}{K(\phi)}\epsilon^{ijk}\d_k \chi
\ee
leads to the dual Lagrangian
\be
{\cal L}^{\rm dual}_{\phi \chi}=-\frac{1}{2}(\d_i \phi)^2-V(\phi)-\frac{1}{2K(\phi)}(\d_i \chi)^2\;,
\ee
where again we have inverted the coupling squared, $K(\phi)\rightarrow 1/K(\phi)$.

Introducing a Chern-Simons term with level $\Theta$, and imposing instead that $f_{ij}=\d_i a_j-\d_j a_i$ with Lagrange multiplier 
${\cal A}_i$, times a constant $C$ to be determined, leads to the dual Lagrangian
\be
{\cal L}^{\rm dual}_{\phi {\cal A}_i}=-\frac{1}{2}(\d_i \phi)^2-V(\phi)+\left(\frac{4\pi}{\Theta}\right)^2C^2
\left[-\frac{1}{4}K(\phi)\tilde f_{jk}^2-\frac{\Theta}{4\pi}\epsilon^{ijk}{\cal A}_i \tilde f_{jk}\right]\;,
\ee
where $\tilde f_{ij}=\d_i {\cal A}_j-\d_j {\cal A}_i$. With the choice
\be
C=\frac{\frac{\Theta}{4\pi}}{\sqrt{K^2+\frac{\Theta^2}{\pi^2}}}\;,
\ee
the duality transformation is 
\bea
K&\rightarrow& K'=\frac{K}{K^2+\frac{\Theta^2}{\pi^2}}\;,\cr
\frac{\Theta}{\pi}&\rightarrow & \frac{\Theta'}{\pi}=\frac{\frac{\Theta}{\pi}}{K^2+\frac{\Theta^2}{\pi^2}}.
\eea

It is known that the Chern-Simons action can be used as a topological response action in the quantum Hall effect, since we have 
\be
j^i=\sigma_H \epsilon^{ijk}\d_j a_k\equiv \frac{\delta S_{\rm CS}}{\delta a_i}.
\ee

But one can also include the Maxwell term, via 
\be
j^a=\frac{\delta S}{\delta a_a}=K \d_j f^{ja}-\frac{\Theta}{2\pi}\epsilon^{aij}\d_i a_j\simeq 
\frac{K}{\tau}\langle E_a\rangle -\frac{\Theta}{2\pi}\epsilon^{ab}
E_b\;,
\ee
so that $K/\tau$ acts as normal conductivity
$\sigma_{ij,||}=\sigma_{||}\delta_{ij}$, while $\Theta/(2\pi)$ acts as Hall conductivity $\sigma_{ij,\perp}=\sigma_H\epsilon_{ij}$.

Then in terms of the complex conductivity $\sigma=\sigma_{xy}+i\sigma_{xx}$, where $\sigma_H=\sigma_{xy}$ and $\sigma_{||}=
\sigma_{xx}$, the particle-vortex duality action on the conductivity takes the form used in the text, and also found by 
\cite{Burgess:2000kj},
\be
\tilde \sigma=-\frac{1}{\sigma}.
\ee

\section{Review of the calculation of the conductivities/susceptibilities in the 
generalized SYK model}
\label{SYKApp}

This section summarizes the calculation of the transport correlators 
given in~\cite{Davison:2016ngz}.

Start from effective action~(\ref{EffAction}) and expand it up to quadratic 
order in fluctuations $f(\tau)=\tau+\epsilon(\tau)$  
and $\delta\phi=\phi$, $\tilde\phi(\tau)=\phi(\tau)+2\pi i{\cal E}T\epsilon(\tau)$. Fourier transforming the result one obtains
\be
S  \ = \ \frac{NKT}{2} \sum\limits_{\omega_n}\omega_n^2|\tilde\phi(\omega_n)|^2 + \frac{T\gamma}{8\pi^2}\sum\limits_{\omega_n}
\omega_n^2(\omega_n^2-4\pi^2T^2)|\epsilon(\omega_n)|^2.
\ee

From this expression one can compute correlation functions
\be
\langle \tilde\phi(\tau) \tilde\phi(0) \rangle \ = \  \frac{T}{NK}\sum\limits_{\omega_n}\frac{e^{i\omega_n\tau}}{\omega_n^2} \ 
= \ \frac{1}{NKT}\left[\frac{1}{2}\left(T\tau-\frac{1}{2}\right)^2-\frac{1}{24}\right]\;,
\ee

\begin{multline}
\langle \epsilon(\tau) \epsilon(0) \rangle \ = \  \frac{4\pi^2T}{\gamma N}
\sum\limits_{\omega_n}\frac{e^{i\omega_n\tau}}{\omega_n^2(\omega_n^2-4\pi^2T^2)} \\
 = \ \frac{1}{\gamma N T^3}\left[\frac{1}{4\pi^2}+\frac{1}{24}-\frac{1}{2}
 \left(T\tau-\frac{1}{2}\right)^2 + \frac{5}{8\pi^2}\cos(2\pi T\tau) + \frac{1}{2\pi}
 \left(T\tau - \frac{1}{2}\right)\sin(2\pi T\tau) \right].
\end{multline}

What one really wants to know is the correlation functions
\be
\langle \delta Q(\tau) \delta Q(0)\rangle \qquad \text{and} \qquad \langle(\delta E(\tau) - \mu \delta Q(\tau))(\delta E(0) - \mu \delta Q(0))\rangle\,,
\ee
which define the charge and the heat condutivities respectively. Here the charge and heat response "currents" are defined as
\begin{eqnarray}
\delta Q & \equiv & \frac{i}{N}\frac{\delta S}{\delta \phi'} \ = \ i K\tilde{\phi}' \,, \label{dQ}\\ 
 \delta E - \mu_0\delta Q & \equiv & \frac{1}{N}\frac{\delta S}{\delta\epsilon'} \ = \  - \frac{\gamma}{4\pi^2}\left(\epsilon'''(\tau)
 +4\pi^2T^2\epsilon'(\tau)\right) + i2\pi K{\cal E}T\tilde{\phi}'(\tau)\,.
\end{eqnarray}

From the above formulas for the correlators of $\tilde{\phi}$ and $\epsilon$ 
one easily finds the expressions for the correlators of the currents
\begin{eqnarray}
\langle \delta Q(\tau) \delta Q(0)\rangle & = &    \frac{KT}{N} \ = \  \frac{\sigma T}{N}\,, \\
\qquad \langle(\delta E(\tau) - \mu \delta Q(\tau))^2\rangle 
& = & \frac{T^3 \left(\gamma +4 \pi ^2 {\cal E}^2 K\right)}{N} \ = \ \frac{\bar{k}T^3}{N}.
\end{eqnarray}

In~\cite{Davison:2016ngz} $\sigma$ and $k=\bar{k}-4\pi^2{\cal E}^2KT$ are referred as to susceptibilities, 
because they can also be computed as derivatives of the generating function, \emph{e.g.}
\begin{eqnarray}
\sigma & = & - \frac{NT}{Z}\frac{\partial^2 Z}{\partial\mu^2}  \ \longrightarrow \ -N T\frac{\partial^2 S}{\partial\mu^2} \ = \  K\,, \\
\bar{k} & = &   \frac{N}{T^3}\frac{1}{Z}\frac{\partial^2 Z}{\partial\beta^2}  
\ \longrightarrow \ -  N\frac{\partial^2(TS)}{\partial T^2} \ = \ \gamma + 4 \pi ^2 {\cal E}^2 K.
\end{eqnarray}

The arrow indicates the result valid in the large $N$ limit, in which the 
partition function is replaced by the classical action for the perturbations. 
One should also assume that the chemical potential $\mu \sim N\phi'$, 
as for example in equation~(\ref{dQ}). One should also use 
a similar relation between $T$ and $\epsilon'$ to derive the last relation 
of the second line.

%%%%%%%%%%%%%%%%%%%%%%%%%%%%%%%%%%%%%%%%%%%%%%%%%%%%%%%%%%%%%%%%%%%%%%%%%%%%%%%%%%%%%%%%
\bibliography{dualityWFSYK}

\providecommand{\href}[2]{#2}\begingroup\raggedright\begin{thebibliography}{10}

\bibitem{Kane:1997fda}
C.~Kane and M.~P. Fisher, ``{Quantized thermal transport in the fractional
  quantum Hall effect},''
  \href{http://dx.doi.org/10.1103/PhysRevB.55.15832}{{\em Phys. Rev. B} {\bf
  55} (1997) no.~23, 15832--15837},
  \href{http://arxiv.org/abs/cond-mat/9603118}{{\tt arXiv:cond-mat/9603118}}.

\bibitem{Read:1999fn}
N.~Read and D.~Green, ``{Paired states of fermions in two-dimensions with
  breaking of parity and time reversal symmetries, and the fractional quantum
  Hall effect},'' \href{http://dx.doi.org/10.1103/PhysRevB.61.10267}{{\em Phys.
  Rev. B} {\bf 61} (2000)  10267},
  \href{http://arxiv.org/abs/cond-mat/9906453}{{\tt arXiv:cond-mat/9906453}}.

\bibitem{Witten:2003ya}
E.~Witten, ``{SL(2,Z) action on three-dimensional conformal field theories with
  Abelian symmetry},'' \href{http://arxiv.org/abs/hep-th/0307041}{{\tt
  arXiv:hep-th/0307041}}.

\bibitem{Burgess:2000kj}
C.~Burgess and B.~P. Dolan, ``{Particle vortex duality and the modular group:
  Applications to the quantum Hall effect and other 2-D systems},''
  \href{http://dx.doi.org/10.1103/PhysRevB.63.155309}{{\em Phys. Rev. B} {\bf
  63} (2001)  155309}, \href{http://arxiv.org/abs/hep-th/0010246}{{\tt
  arXiv:hep-th/0010246}}.

\bibitem{Murugan:2014sfa}
J.~Murugan, H.~Nastase, N.~Rughoonauth, and J.~P. Shock, ``{Particle-vortex and
  Maxwell duality in the $AdS_4\times \mathbb{CP}^3$/ABJM correspondence},''
  \href{http://dx.doi.org/10.1007/JHEP10(2014)051}{{\em JHEP} {\bf 10} (2014)
  051}, \href{http://arxiv.org/abs/1404.5926}{{\tt arXiv:1404.5926 [hep-th]}}.

\bibitem{Nastase:2017cxp}
H.~Nastase, {\em {String Theory Methods for Condensed Matter Physics}}.
\newblock Cambridge University Press,
2017.
\newblock
%%CITATION = INSPIRE-1638019;%%.

\bibitem{Hartnoll:2007ai}
S.~A. Hartnoll and P.~Kovtun, ``{Hall conductivity from dyonic black holes},''
  \href{http://dx.doi.org/10.1103/PhysRevD.76.066001}{{\em Phys. Rev. D} {\bf
  76} (2007)  066001}, \href{http://arxiv.org/abs/0704.1160}{{\tt
  arXiv:0704.1160 [hep-th]}}.

\bibitem{Hartnoll:2007ih}
S.~A. Hartnoll, P.~K. Kovtun, M.~Muller, and S.~Sachdev, ``{Theory of the
  Nernst effect near quantum phase transitions in condensed matter, and in
  dyonic black holes},''
  \href{http://dx.doi.org/10.1103/PhysRevB.76.144502}{{\em Phys. Rev. B} {\bf
  76} (2007)  144502}, \href{http://arxiv.org/abs/0706.3215}{{\tt
  arXiv:0706.3215 [cond-mat.str-el]}}.

\bibitem{Alejo:2019utd}
L.~Alejo, P.~Goulart, and H.~Nastase, ``{S-duality, entropy function and
  transport in $AdS_4/CMT_3$},''
  \href{http://dx.doi.org/10.1007/JHEP09(2019)003}{{\em JHEP} {\bf 09} (2019)
  003}, \href{http://arxiv.org/abs/1905.04898}{{\tt arXiv:1905.04898
  [hep-th]}}.

\bibitem{Alejo:2019hnb}
L.~Alejo and H.~Nastase, ``{Particle-vortex duality and theta terms in AdS/CMT
  applications},'' \href{http://dx.doi.org/10.1007/JHEP08(2019)095}{{\em JHEP}
  {\bf 08} (2019)  095}, \href{http://arxiv.org/abs/1905.03549}{{\tt
  arXiv:1905.03549 [hep-th]}}.

\bibitem{Blake:2015ina}
M.~Blake, A.~Donos, and N.~Lohitsiri, ``{Magnetothermoelectric Response from
  Holography},'' \href{http://dx.doi.org/10.1007/JHEP08(2015)124}{{\em JHEP}
  {\bf 08} (2015)  124}, \href{http://arxiv.org/abs/1502.03789}{{\tt
  arXiv:1502.03789 [hep-th]}}.

\bibitem{Herzog:2007ij}
C.~P. Herzog, P.~Kovtun, S.~Sachdev, and D.~T. Son, ``{Quantum critical
  transport, duality, and M-theory},''
  \href{http://dx.doi.org/10.1103/PhysRevD.75.085020}{{\em Phys. Rev. D} {\bf
  75} (2007)  085020}, \href{http://arxiv.org/abs/hep-th/0701036}{{\tt
  arXiv:hep-th/0701036}}.

\bibitem{Davison:2016ngz}
R.~A. Davison, W.~Fu, A.~Georges, Y.~Gu, K.~Jensen, and S.~Sachdev,
  ``{Thermoelectric transport in disordered metals without quasiparticles: The
  Sachdev-Ye-Kitaev models and holography},''
  \href{http://dx.doi.org/10.1103/PhysRevB.95.155131}{{\em Phys. Rev. B} {\bf
  95} (2017) no.~15, 155131}, \href{http://arxiv.org/abs/1612.00849}{{\tt
  arXiv:1612.00849 [cond-mat.str-el]}}.

\bibitem{Kadanoff:1963}
L.~P. Kadanoff and P.~C. Martin, ``Hydrodynamic equations and correlation
  functions,'' \href{http://dx.doi.org/10.1016/0003-4916(63)90078-2}{{\em
  Annals of Physics} {\bf 24} (1963)  419 -- 469}.
  \url{http://www.sciencedirect.com/science/article/pii/0003491663900782}.

\bibitem{Wen:1992uk}
X.~Wen and A.~Zee, ``{A Classification of Abelian quantum Hall states and
  matrix formulation of topological fluids},''
  \href{http://dx.doi.org/10.1103/PhysRevB.46.2290}{{\em Phys. Rev. B} {\bf 46}
  (1992)  2290--2301}.

\bibitem{Bloete:1986qm}
H.~Bloete, J.~L. Cardy, and M.~Nightingale, ``{Conformal Invariance, the
  Central Charge, and Universal Finite Size Amplitudes at Criticality},''
  \href{http://dx.doi.org/10.1103/PhysRevLett.56.742}{{\em Phys. Rev. Lett.}
  {\bf 56} (1986)  742--745}.

\bibitem{Affleck:1986bv}
I.~Affleck, ``{Universal Term in the Free Energy at a Critical Point and the
  Conformal Anomaly},''
  \href{http://dx.doi.org/10.1103/PhysRevLett.56.746}{{\em Phys. Rev. Lett.}
  {\bf 56} (1986)  746--748}.

\bibitem{Cappelli:2001mp}
A.~Cappelli, M.~Huerta, and G.~R. Zemba, ``{Thermal transport in chiral
  conformal theories and hierarchical quantum Hall states},''
  \href{http://dx.doi.org/10.1016/S0550-3213(02)00340-1}{{\em Nucl. Phys. B}
  {\bf 636} (2002)  568--582},
  \href{http://arxiv.org/abs/cond-mat/0111437}{{\tt arXiv:cond-mat/0111437}}.

\bibitem{Senthil:1998qu}
T.~Senthil, M.~P. Fisher, L.~Balents, and C.~Nayak, ``Quasiparticle transport
  and localization in high-t c superconductors,'' {\em Phys. Rev. Lett.} {\bf
  81} (1998) no.~21, 4704.

\bibitem{Donos:2017mhp}
A.~Donos, J.~P. Gauntlett, T.~Griffin, N.~Lohitsiri, and L.~Melgar,
  ``{Holographic DC conductivity and Onsager relations},''
  \href{http://dx.doi.org/10.1007/JHEP07(2017)006}{{\em JHEP} {\bf 07} (2017)
  006}, \href{http://arxiv.org/abs/1704.05141}{{\tt arXiv:1704.05141
  [hep-th]}}.

\bibitem{Iqbal:2008by}
N.~Iqbal and H.~Liu, ``{Universality of the hydrodynamic limit in AdS/CFT and
  the membrane paradigm},''
  \href{http://dx.doi.org/10.1103/PhysRevD.79.025023}{{\em Phys. Rev. D} {\bf
  79} (2009)  025023}, \href{http://arxiv.org/abs/0809.3808}{{\tt
  arXiv:0809.3808 [hep-th]}}.

\bibitem{Donos:2015bxe}
A.~Donos, J.~P. Gauntlett, T.~Griffin, and L.~Melgar, ``{DC Conductivity of
  Magnetised Holographic Matter},''
  \href{http://dx.doi.org/10.1007/JHEP01(2016)113}{{\em JHEP} {\bf 01} (2016)
  113}, \href{http://arxiv.org/abs/1511.00713}{{\tt arXiv:1511.00713
  [hep-th]}}.

\bibitem{Erdmenger:2016wyp}
J.~Erdmenger, D.~Fernandez, P.~Goulart, and P.~Witkowski, ``{Conductivities
  from attractors},'' \href{http://dx.doi.org/10.1007/JHEP03(2017)147}{{\em
  JHEP} {\bf 03} (2017)  147}, \href{http://arxiv.org/abs/1611.09381}{{\tt
  arXiv:1611.09381 [hep-th]}}.

\bibitem{Melnikov:2012tb}
D.~Melnikov, E.~Orazi, and P.~Sodano, ``{On the AdS/BCFT Approach to Quantum
  Hall Systems},'' \href{http://dx.doi.org/10.1007/JHEP05(2013)116}{{\em JHEP}
  {\bf 05} (2013)  116}, \href{http://arxiv.org/abs/1211.1416}{{\tt
  arXiv:1211.1416 [hep-th]}}.

\bibitem{Melnikov:2017wfg}
D.~Melnikov, ``{Topological transport from a black hole},''
  \href{http://dx.doi.org/10.1016/j.physletb.2018.01.027}{{\em Phys. Lett. B}
  {\bf 778} (2018)  174--177}, \href{http://arxiv.org/abs/1704.03973}{{\tt
  arXiv:1704.03973 [hep-th]}}.

\bibitem{Shapere:1988zv}
A.~D. Shapere and F.~Wilczek, ``{Selfdual Models with Theta Terms},''
  \href{http://dx.doi.org/10.1016/0550-3213(89)90016-3}{{\em Nucl. Phys. B}
  {\bf 320} (1989)  669--695}.

\bibitem{Girvin:1984zz}
S.~Girvin, ``{Particle-hole symmetry in the anomalous quantum Hall effect},''
  \href{http://dx.doi.org/10.1103/PhysRevB.29.6012}{{\em Phys. Rev. B} {\bf 29}
  (1984)  6012--6014}.

\bibitem{Jain:1990sc}
J.~K. Jain, S.~Kivelson, and N.~Trivedi, ``Scaling theory of the fractional
  quantum hall effect,'' {\em Phys. Rev. Lett.} {\bf 64} (1990) no.~11, 1297.

\bibitem{Jain:1992hi}
J.~K. Jain and V.~Goldman, ``Hierarchy of states in the fractional quantum hall
  effect,'' {\em Phys. Rev. B} {\bf 45} (1992) no.~3, 1255.

\bibitem{Fradkin:1996xb}
E.~H. Fradkin and S.~Kivelson, ``{Modular invariance, selfduality and the phase
  transition between quantum Hall plateaus},''
  \href{http://dx.doi.org/10.1016/0550-3213(96)00310-0}{{\em Nucl. Phys. B}
  {\bf 474} (1996)  543--574},
  \href{http://arxiv.org/abs/cond-mat/9603156}{{\tt arXiv:cond-mat/9603156}}.

\bibitem{Evers:2008zz}
F.~Evers and A.~D. Mirlin, ``{Anderson transitions},''
  \href{http://dx.doi.org/10.1103/RevModPhys.80.1355}{{\em Rev. Mod. Phys.}
  {\bf 80} (2008)  1355--1417},
\href{http://arxiv.org/abs/0707.4378}{{\tt arXiv:0707.4378
  [cond-mat.mes-hall]}}.
%%CITATION = ARXIV:0707.4378;%%.

\bibitem{Amoretti:2020mkp}
A.~Amoretti, D.~K. Brattan, N.~Magnoli, and M.~Scanavino, ``{Magneto-thermal
  transport implies an incoherent Hall conductivity},''
  \href{http://dx.doi.org/10.1007/JHEP08(2020)097}{{\em JHEP} {\bf 08} (2020)
  097}, \href{http://arxiv.org/abs/2005.09662}{{\tt arXiv:2005.09662
  [hep-th]}}.

\bibitem{Itzhaki:1998dd}
N.~Itzhaki, J.~M. Maldacena, J.~Sonnenschein, and S.~Yankielowicz,
  ``{Supergravity and the large N limit of theories with sixteen
  supercharges},'' \href{http://dx.doi.org/10.1103/PhysRevD.58.046004}{{\em
  Phys. Rev. D} {\bf 58} (1998)  046004},
  \href{http://arxiv.org/abs/hep-th/9802042}{{\tt arXiv:hep-th/9802042}}.

\bibitem{Aharony:2008ug}
O.~Aharony, O.~Bergman, D.~L. Jafferis, and J.~Maldacena, ``{N=6 superconformal
  Chern-Simons-matter theories, M2-branes and their gravity duals},''
  \href{http://dx.doi.org/10.1088/1126-6708/2008/10/091}{{\em JHEP} {\bf 10}
  (2008)  091}, \href{http://arxiv.org/abs/0806.1218}{{\tt arXiv:0806.1218
  [hep-th]}}.

\bibitem{Sachdev:2019bjn}
S.~Sachdev, ``{Universal low temperature theory of charged black holes with
  AdS$_2$ horizons},'' \href{http://dx.doi.org/10.1063/1.5092726}{{\em J. Math.
  Phys.} {\bf 60} (2019) no.~5, 052303},
  \href{http://arxiv.org/abs/1902.04078}{{\tt arXiv:1902.04078 [hep-th]}}.

\bibitem{Kitaev:2015}
A.~Y. Kitaev, ``Entanglementin strongly-correlated quantum matter.'' KITP talk,
  University of California, Santa Barbara, 2015.

\bibitem{Maldacena:2016hyu}
J.~Maldacena and D.~Stanford, ``{Remarks on the Sachdev-Ye-Kitaev model},''
  \href{http://dx.doi.org/10.1103/PhysRevD.94.106002}{{\em Phys. Rev.} {\bf
  D94} (2016) no.~10, 106002},
\href{http://arxiv.org/abs/1604.07818}{{\tt arXiv:1604.07818 [hep-th]}}.
%%CITATION = ARXIV:1604.07818;%%.

\bibitem{Murugan:2016zal}
J.~Murugan and H.~Nastase, ``{Particle-vortex duality in topological insulators
  and superconductors},'' \href{http://dx.doi.org/10.1007/JHEP05(2017)159}{{\em
  JHEP} {\bf 05} (2017)  159}, \href{http://arxiv.org/abs/1606.01912}{{\tt
  arXiv:1606.01912 [hep-th]}}.

\bibitem{peterson2010kelvin}
M.~R. Peterson and B.~S. Shastry, ``Kelvin formula for thermopower,'' {\em
  Physical Review B} {\bf 82} (2010) no.~19, 195105.

\end{thebibliography}\endgroup
\bibliographystyle{utphys}
%%%%%%%%%%%%%%%%%%%%%%%%%%%%%%%%%%%%%%%%%%%%%%%%%%%%%%%%%%%%%%%%%%%%%%%%%%%%%%%%%%%%%%%%

\end{document}